\documentclass[
preprintnumbers,showpacs,amsmath,amssymb,aps,prd]{revtex4}
\usepackage{dcolumn}
\usepackage{amsthm,amscd,mathrsfs}
\usepackage{amsfonts,bm,amssymb,latexsym}

\def\be{\begin{equation}}
\def\ee{\end{equation}}
\def\bea{\begin{eqnarray}}
\def\eea{\end{eqnarray}}
\def\ba{\begin{array}}
\def\ea{\end{array}}
\def\nn{\nonumber}
\def\p{\partial}
\def\cD{\mathcal{D}}
\def\cL{\mathcal{L}}

\begin{document}

\preprint{arXiv:0906.2049v4 [hep-th]}

\title{Separability of massive field equations for spin-0 and spin-$1/2$
charged particles in the general non-extremal rotating charged black holes
in minimal five-dimensional gauged supergravity}

\author{Shuang-Qing Wu\footnote{sqwu@phy.ccnu.edu.cn}}
\affiliation{College of Physical Science and Technology, Central China Normal University,
Wuhan, Hubei 430079, People's Republic of China}

\begin{abstract}
We continue to investigate the separability of massive field equations for spin-0
and spin-$1/2$ charged particles in the general, non-extremal, rotating, charged,
Chong-Cveti\v{c}-L\"{u}-Pope black holes with two independent angular momenta and
a non-zero cosmological constant in minimal $D = 5$ gauged supergravity theory. We
show that the complex Klein-Gordon equation and the modified Dirac equation with the
inclusion of an extra counter-term can be separated by variables into purely radial
and purely angular parts in this general Einstein-Maxwell-Chern-Simons background
spacetime. A second-order symmetry operator that commutes with the complex Laplacian
operator is constructed from the separated solutions and expressed compactly in terms
of a rank-2 St\"{a}ckel-Killing tensor which admits a simple diagonal form in the
chosen pentad one-forms so that it can be understood as the square of a rank-3 totally
anti-symmetric tensor. A first-order symmetry operator that commutes with the modified
Dirac operator is expressed in terms of a rank-3 generalized Killing-Yano tensor and
its covariant derivative. The Hodge dual of this generalized Killing-Yano tensor is a
generalized principal conformal Killing-Yano tensor of rank-2, which can generate a
``tower'' of generalized (conformal) Killing-Yano and St\"{a}ckel-Killing tensors that
are responsible for the whole hidden symmetries of this general, rotating, charged,
Kerr-AdS black hole geometry. In addition, the first laws of black hole thermodynamics
have been generalized to the case that the cosmological constant can be viewed as a
thermodynamical variable.
\end{abstract}

\maketitle

\section{Introduction}

In 1976, Chandrasekhar \cite{SC} showed that the massive Dirac's equation is separable
in the Kerr geometry using the Newman-Penrose's null-tetrad formalism. Subsequently,
this work was further extended by Page \cite{DNP} and Lee \cite{CHL} to the case of
a Kerr-Newman black hole. Later on, Carter and McLenaghan \cite{CM} found that the
separability of the Dirac equation in the Kerr geometry, is related to the fact that
the skew-symmetric tensor corresponding to the two-index Killing spinor admitted by
the Kerr metric is a rank-2 anti-symmetric Killing-Yano tensor, and its square is
just a second-order symmetric St\"{a}ckel-Killing tensor discovered by Carter \cite{BC1}.
What is more, using the Killing-Yano tensor, they had constructed a first-order differential
operator that commutes with the Dirac operator. The separation constant appearing in
the separated solutions to the Dirac equation acts as the eigenvalue of this first
order symmetry operator. The essential property that allows the construction of such
a symmetry operator is the existence of a Killing-Yano tensor in the Kerr spacetime.

Recently, we \cite{WuMP,WuKAdS} have investigated the separability of Dirac's equation
and its relation to the Killing-Yano tensor of rank-3 in the five-dimensional Myers-Perry
\cite{MP} and Kerr-(anti-)de Sitter \cite{HHT} black hole spacetime with two unequal
angular momenta. A first-order symmetry operator commuting with the Dirac operator has
been constructed by using the rank-3 Killing-Yano tensor whose square is just the rank-2
symmetric St\"{a}ckel-Killing tensor. In addition, we have constructed a second-order
symmetry operator that commutes with the scalar Laplacian operator.

In a subsequent paper \cite{WuCY}, we have extended this work to investigate the separability
of a massive fermion field equation for spin-$1/2$ spinor particles in a five-dimensional
rotating, charged, Cveti\v{c}-Youm \cite{CY} black hole with two different angular momenta
and three equal charges. This black hole solution represents a natural generalization
of the four-dimensional Kerr-Newman solution to five dimensions with the inclusion
of a Chern-Simons term to the Maxwell equation, and belongs to the classes of
Einstein-Mawxell-Chern-Simons (EMCS) black holes within minimal $D = 5$ ungauged
supergravity theory. It has been shown that the usual massive Dirac equation can not
be separated by variables into purely radial and purely angular parts in this general
rotating, charged black hole spacetime with two independent angular momenta. However,
if an additional counter-term is supplemented into the usual Dirac operator, then the
modified Dirac field equation for the spin-$1/2$ spinor particles is separable in this
five-dimensional rotating charged black hole background geometry. A first-order symmetry
operator that commutes with the modified Dirac operator has exactly the same form as that
previously found in the uncharged Myers-Perry black hole case. This operator is expressed
in terms of a rank-3 totally anti-symmetric tensor and its covariant derivative. The
anti-symmetric tensor obeys a generalized Killing-Yano equation and its square is a
second-order symmetric St\"{a}ckel-Killing tensor admitted by the five-dimensional
rotating charged black hole spacetime. Furthermore, the inclusion of such an additional
counter-term can be geometrically understood \cite{KKY} as a natural consequence if
one identifies the dual Maxwell three-form with a generalized ``torsion''. This ``formal''
identification has many appealing features, for example, the Maxwell-Chern-Simons
equations can be ``simplified'' to the standard Maxwell equations, and the generalized
(conformal) Killing-Yano tensors ``possess'' many of the properties of the standard ones.
However, it will unavoidably complicate the Einstein's gravitational part since the
inclusion of a torsion means that one has to deal with an asymmetric connection,
therefore it is not necessary to make such a geometric identification.

In this paper, we will demonstrate that our previous analysis done in \cite{WuCY} is
directly applicable to deal with the case of a nonzero cosmological constant, namely,
the five-dimensional Chong-Cveti\v{c}-L\"{u}-Pope \cite{CCLP} (alternatively, EMCS-Kerr-anti
de Sitter) black hole. More specifically, we shall study the separation of variables
for the massive field equations of spin$-0$ and spin-$1/2$ charged particles in this
five-dimensional, general, non-extremal, rotating, charged EMCS-Kerr-AdS black holes
with two independent angular momenta and a negative cosmological constant. This black
hole metric is an exact solution within the framework of minimal $D = 5$ gauged supergravity
theory, and has attracted a lot of interest \cite{DKL,ANA1,ANA2,EMCSbht,EBHCFT} since
its being constructed. It has been shown \cite{DKL,ANA1} that the geodesic equation, the
Hamilton-Jacobi equation, the Klein-Gordon equation, and the stationary string in this
spacetime are all separable since Frolov and his collaborators \cite{FKKF} initiated
the studies (see \cite{VFDK} for a review and references therein) of hidden symmetries
and separability properties of higher-dimensional rotating black hole spacetimes. The
separability properties of these equations are also intimately associated with the
existence of a second-order symmetric St\"{a}ckel-Killing tensor. In particular, the
authors of Ref. \cite{DKL} have demonstrated that the usual Dirac equation allows the
separation of variables only in the special case with two equal magnitude angular momenta.

Therefore, the question of the separability of a spin-$1/2$ field equation and its relation
to the rank-3 antisymmetric Killing-Yano tensor in the general Chong-Cveti\v{c}-L\"{u}-Pope
black hole background spacetime remains unsolved in a satisfactory manner similar to the
uncharged case before we briefly announced in the conclusion section of \cite{WuCY} that
the whole hidden symmetries of this general EMCS-Kerr-AdS black hole geometry are controlled
by a rank-3 generalized Killing-Yano tensor and its Hodge dual two-form --- a generalized
principal conformal Killing-Yano tensor of rank-2. To resolve this question constitutes
one of the main subjects of this paper. Specifically speaking, we will study the separation
of variables for massive field equations of spin-0 and spin-$1/2$ charged particles in
the general, non-extremal, rotating, charged, Chong-Cveti\v{c}-L\"{u}-Pope black holes
with two independent angular momenta and a non-zero cosmological constant. What is more,
we shall present two symmetry operators that commutes respectively with the complex scalar
Laplacian operator and the modified Dirac operator. These differential operators are,
respectively, expressed in terms of a rank-2 St\"{a}ckel-Killing tensor and its ``square
root'' --- a rank-3 generalized Killing-Yano tensor. They characterize the separation
constants that appear in the separable solutions of the massive Klein-Gordon scalar
field equation and the modified Dirac equation with the minimal gauge coupling term.
The Hodge dual of this generalized Killing-Yano tensor is a generalized principal
conformal Killing-Yano tensor of rank-2, which can generate the ``tower'' of generalized
Killing-Yano and St\"{a}ckel-Killing tensors that are responsible for the whole hidden
symmetries of the general EMCS-Kerr-AdS black hole geometry. The separability properties
of these wave equations are shown to be closely connected with the existence of these
St\"{a}ckel-Killing and generalized (conformal) Killing-Yano tensors admitted by the
five-dimensional EMCS-Kerr-AdS metric.

Our paper is organized as follows. In Sec. \ref{EMCSKAdS}, a simple, elegant form for
the line element of the five-dimensional Chong-Cveti\v{c}-L\"{u}-Pope black hole solution
is presented in the Boyer-Lindquist coordinates, which is very convenient for us to
explicitly construct the local Lorentzian orthonormal coframe one-forms (pentad). We also
present some interesting properties of this spacetime. In particular, we will generalize
the first law of black hole thermodynamics to the case when the cosmological constant is
viewed as a thermodynamical variable. Section \ref{TGCKY} summarizes the main results reported
in this paper. In this section, we will discuss the hidden symmetries of the five-dimensional
Chong-Cveti\v{c}-L\"{u}-Pope black hole and present the explicit expressions of a rank-2
St\"{a}ckel-Killing tensor, a rank-3 generalized Killing-Yano tensor, a rank-2 generalized
conformal Killing-Yano tensor. A second-order symmetry operator that commutes with the
complex scalar Laplacian operator and a first-order differential operator that commutes
with the modified Dirac operator are, respectively, expressed in terms of the rank-2
St\"{a}ckel-Killing tensor and the rank-3 generalized Killing-Yano tensor. Then in Sec.
\ref{KGSK}, we focus on the separation of variables for a massive complex Klein-Gordon
scalar field equation in this spacetime background and use the separated solutions to
directly construct a second-order differential operator that commutes with the complex
scalar Laplacian operator, from which we can extract a concise expression for the rank-2
St\"{a}ckel-Killing tensor. Section \ref{SmDE} is devoted to the separation of variables
for a massive spinor field equation in the general five-dimensional EMCS-Kerr-AdS black
hole geometry. Adopting the f\"{u}nfbein formalism, we will show that the modified Dirac's
equation with a minimal gauge coupling term can be separated into purely radial and
purely angular parts. In Sec. \ref{dEgKY}, we shall demonstrate that the dual first
order differential operator previously constructed in terms of the rank-3 generalized
Killing-Yano tensor (and its covariant derivative) has an eigenvalue as the separation
constant appearing in the separated parts of the modified Dirac's equation, which means
that this first-order symmetry operator commutes with the modified Dirac operator. The
last section \ref{CoRe} ends up with a summary of our work done in this paper. In the
appendix, we give various useful expressions for the details of our calculations. The
spin-connection one-forms are calculated by the first Cartan's structure equation from
the exterior differential of the pentad. The corresponding spinor-connection one-forms,
the Riemann curvature two-forms, and the Weyl curvature two-forms are also given in this
pentad formalism.

\section{Chong-Cveti\v{c}-L\"{u}-Pope black hole solution and
its some fundamental properties} \label{EMCSKAdS}

Although the neutrally-charged generalizations of the Kerr metric to higher dimensions
were obtained \cite{MP} many years ago, higher-dimensional charged generalizations of
the four-dimensional Kerr-Newman black hole still remain unknown in pure Einstein-Maxwell
theory. In the simplest $D = 5$ case, the inclusion of a Chern-Simons term to the Maxwell
equation makes it easier to solve the field equations in the minimal supergravity theory.
Up to now, almost all exact solutions known for rotating charged black holes in five
dimensions fall into the framework of EMCS supergravity theory.

The bosonic part of minimal $D = 5$ gauged supergravity theory consists of the metric and
a one-form gauge field, and is given by
\be
S = \frac{1}{16\pi}\int d^5x \Big[\sqrt{-g}\big(R +12/l^2 -F_{\mu\nu}F^{\mu\nu}\big)
 -\frac{2}{3\sqrt{3}}\epsilon^{\mu\nu\alpha\beta\gamma}F_{\mu\nu}
 F_{\alpha\beta}\mathcal{A}_{\gamma}\Big] \, .
\ee
Variation of this action yields the Einstein equation and the Maxwell-Chern-Simons equation
\bea
&& R_{\mu\nu} -\frac{1}{2}g_{\mu\nu}R -\frac{6}{l^2}g_{\mu\nu}
  = 2T_{\mu\nu} \equiv 2\Big(F_{\mu\alpha}F_{\nu}^{~\alpha}
  -\frac{1}{4}g_{\mu\nu}F_{\alpha\beta}F^{\alpha\beta}\Big) \, , \qquad \\
&& \p_{\nu}\Big(\sqrt{-g}F^{\mu\nu} +\frac{1}{\sqrt{3}}
  \epsilon^{\mu\nu\alpha\beta\gamma} \mathcal{A}_{\alpha}F_{\beta\gamma}\Big) = 0 \, .
\eea
A general solution that describes a five-dimensional non-extremal rotating, charged black
hole with two independent angular momenta and a negative cosmological constant was constructed
in Ref. \cite{CCLP}. The metric and the gauge potential given below simultaneously solve
the Einstein equation and the Maxwell-Chern-Simons equation. We present here an elegant
expression for the line element that had already been obtained by the present author soon
after the announcement of this solution in the e-print archive.

For our purpose in this paper, we rewrite the metric for the five-dimensional rotating
charged EMCS-Kerr-AdS black hole solution into a quasi-diagonal form so that we can easily
construct a local Lorentzian orthonormal pentad with which the spinor field equation for
the spin-$1/2$ particles can be decoupled into purely radial and purely angular parts.
As did in Ref. \cite{WuCY}, we find that the line element can be recast into a simple
form in terms of the Boyer-Lindquist coordinates as follows:
\bea
ds^2 &=& g_{\mu\nu}dx^{\mu}dx^{\nu} \nn \\
 &=& -\frac{\Delta_r}{\Sigma} X^2 +\frac{\Sigma}{\Delta_r}dr^2
 +\frac{\Sigma}{\Delta_{\theta}}d\theta^2
 +\frac{\Delta_{\theta}(a^2-b^2)^2\sin^2\theta\cos^2\theta}{p^2\Sigma} Y^2
 +\Big(\frac{ab}{rp} Z +\frac{Qp}{r\Sigma} X\Big)^2 \, ,
\label{EMCSKADS}
\eea
and the gauge potential is
\be
\mathcal{A} = \frac{\sqrt{3}Q}{2\Sigma} X \, ,
\ee
where we denote
\begin{subequations}\begin{align}
X &= dt -\frac{a\sin^2\theta}{\chi_a}d\phi
 -\frac{b\cos^2\theta}{\chi_b}d\psi \, , \\
Y &= dt -\frac{(r^2+a^2)a}{(a^2-b^2)\chi_a}d\phi
 -\frac{(r^2+b^2)b}{(b^2-a^2)\chi_b}d\psi \, , \\
Z &= dt -\frac{(r^2+a^2)\sin^2\theta}{a\chi_a}d\phi
 -\frac{(r^2+b^2)\cos^2\theta}{b\chi_b}d\psi \, ,
\end{align}\end{subequations}
and
\bea
&& \Delta_r = (r^2+a^2)(r^2+b^2)\Big(\frac{1}{r^2} +\frac{1}{l^2}\Big) -2M
 +\frac{Q^2+2Qab}{r^2} \, , \qquad \Delta_{\theta} = 1 -\frac{p^2}{l^2} \, , \nn \\
&& \Sigma = r^2 +p^2 \, , \qquad p = \sqrt{a^2\cos^2\theta +b^2\sin^2\theta} \, , \qquad
 \chi_a = 1 -\frac{a^2}{l^2} \, , \qquad \chi_b = 1 -\frac{b^2}{l^2} \, . \nn
\eea
Here the parameters ($M, Q, a, b, l$) are related to the mass, two independent angular
momenta of the black hole, and the cosmological constant.

The new form of the five-dimensional EMCS-Kerr-AdS metric (\ref{EMCSKADS}) admits a simple,
diagonal form:
\be
ds^2 = \eta_{AB}e^A\otimes e^B = -(e^0)^2 +(e^1)^2 +(e^2)^2 +(e^3)^2 +(e^5)^2 \, ,
\ee
which allows us to choose the following local Lorentzian basis one-forms (pentad) defined
as $e^A = e^A_{~\mu}dx^{\mu}$ orthonormal with respect to the flat (Lorentzian) metric
$\eta_{AB} = {\rm diag} (-1, 1, 1, 1, 1)$,
\be
 e^0 = \sqrt{\frac{\Delta_r}{\Sigma}} X \, , \quad
 e^1 = \sqrt{\frac{\Sigma}{\Delta_r}}dr \, , \quad
 e^2 = \sqrt{\frac{\Sigma}{\Delta_p}}dp \, , \quad
 e^3 = \sqrt{\frac{\Delta_p}{\Sigma}} Y \, , \quad
 e^5 = -\Big(\frac{ab}{rp} Z +\frac{Qp}{r\Sigma} X\Big) \, .
\label{pentad}
\ee
Throughout this paper, we shall adopt conventions as follows: Greek letters $\mu, \nu$
run over five-dimensional spacetime coordinate indices $\{t, r, \theta, \phi, \psi\}$,
while Latin letters $A, B$ denote local orthonormal (Lorentzian) frame indices $\{0, 1,
2, 3, 5\}$.

The above line element (\ref{EMCSKADS}) is written in a coordinate frame rotating at
infinity. To compute the physical mass, angular momenta and electric charge, one has
to change the metric to a coordinate frame nonrotating at infinity by making the
transformations: $\phi = \widetilde{\phi} -at/l^2$ and $\psi = \widetilde{\psi} -bt/l^2$.

The outer event horizon is determined by the largest root of $\Delta_{r_+} = 0$. The
Hawking temperature $T = \kappa/(2\pi)$ and the Bekenstein-Hawking entropy $S = A/4$
with respect to this horizon can be easily computed as
\bea
T &=& \frac{r_+^4\big[1 +(2r_+^2 +a^2 +b^2)/l^2\big] -(Q +ab)^2}{2\pi
 r_+\big[(r_+^2 +a^2)(r_+^2 +b^2) +Qab\big]} \, , \\
S &=& \pi^2\frac{(r_+^2 +a^2)(r_+^2+b^2) +Qab}{2\chi_a\chi_br_+} \, ,
\eea
while the angular velocities and the electro-static potential, measured relative to a
frame that is non-rotating at infinity, are given by
\bea
\Omega_a &=& \frac{a(r_+^2 +b^2)(1 +r_+^2/l^2) +Qb}{(r_+^2 +a^2)(r_+^2 +b^2) +Qab} \, , \\
\Omega_b &=& \frac{b(r_+^2 +a^2)(1 +r_+^2/l^2) +Qa}{(r_+^2 +a^2)(r_+^2 +b^2) +Qab} \, , \\
\Phi &=& \frac{\sqrt{3}Qr_+^2/2}{(r_+^2 +a^2)(r_+^2+b^2) +Qab} \, .
\eea

The physical mass, two angular momenta, and the electric charge are given by
\cite{CCLP,EMCSbht}
\bea
\mathcal{M} &=& \frac{\pi}{2\chi_a\chi_b}\Big[\Big(M +\frac{Qab}{l^2}\Big)
 \Big(\frac{1}{\chi_a} +\frac{1}{\chi_b}\Big) -\frac{1}{2}M\Big] \, , \\
J_a &=& \frac{\pi\big[2Ma +Qb(2 -\chi_a)\big]}{4\chi_a^2\chi_b} \, , \\
J_b &=& \frac{\pi\big[2Mb +Qa(2 -\chi_b)\big]}{4\chi_a\chi_b^2} \, , \\
\mathcal{Q} &=& \frac{\sqrt{3}\pi Q}{2\chi_a\chi_b} \, ,
\eea
which fulfill the closed forms for the first law of black hole thermodynamics
\begin{subequations}\begin{align}
\frac{2}{3}\mathcal{M} &= TS +\Omega_aJ_a +\Omega_bJ_b
 +\frac{2}{3}\Phi\mathcal{Q} -\frac{1}{3}\Theta l \, , \\
d\mathcal{M} &= TdS +\Omega_adJ_a +\Omega_bdJ_b +\Phi d\mathcal{Q} -\Theta dl \, ,
\end{align}\end{subequations}
where we have introduced the generalized force conjugate to the cosmological radius $l$ as
\bea
\Theta = \frac{\pi}{2\chi_a\chi_b l}\Big[M +\Big(M +\frac{Qab}{l^2}\Big)
 \Big(\frac{1}{\chi_a} +\frac{1}{\chi_b} -\frac{3}{1 +r_+^2/l^2}\Big)
 -\frac{3Q^2}{2l^2(1 +r_+^2/l^2)}\Big] \, .
\eea

In practice, it is very useful to adopt $p$ rather than $\theta$ as the appropriate
angle coordinate. What is more, the radial part and the angular part can be presented in
a symmetric manner. In what follows, we shall adopt $p$ as the convenient angle coordinate
throughout this article. In doing so, the five-dimensional Chong-Cveti\v{c}-L\"{u}-Pope
metric can be rewritten as
\be
ds^2 = -\frac{\Delta_r}{\Sigma} X^2 +\frac{\Sigma}{\Delta_r}dr^2
  +\frac{\Sigma}{\Delta_p}dp^2 +\frac{\Delta_p}{\Sigma} Y^2
  +\Big(\frac{ab}{rp} Z +\frac{Qp}{r\Sigma} X\Big)^2 \, ,
\label{mPDf}
\ee
where
\be
\Delta_p = -(p^2-a^2)(p^2-b^2)\Big(\frac{1}{p^2} -\frac{1}{l^2}\Big) \, ,
\ee
and
\begin{subequations}\begin{align}
X &= dt -\frac{(p^2-a^2)a}{(b^2-a^2)\chi_a}d\phi
 -\frac{(p^2-b^2)b}{(a^2-b^2)\chi_b}d\psi \, , \\
Y &= dt +\frac{(r^2+a^2)a}{(b^2-a^2)\chi_a}d\phi
 +\frac{(r^2+b^2)b}{(a^2-b^2)\chi_b}d\psi\, , \\
Z &= dt -\frac{(r^2+a^2)(p^2-a^2)}{(b^2-a^2)a\chi_a}d\phi
 -\frac{(r^2+b^2)(p^2-b^2)}{(a^2-b^2)b\chi_b}d\psi\, .
\end{align}\end{subequations}

After doing the following coordinate transformations:
\be
t = \tau +(a^2+b^2)u +a^2b^2v \, , \qquad\quad \phi = a\chi_a(u +b^2v) \, , \qquad\quad
\psi = b\chi_b(u +a^2v) \, ,
\ee
we get
\bea
X = d\tau +p^2du \, , \qquad\quad Y = d\tau -r^2du \, , \qquad\quad
Z = d\tau +(p^2-r^2)du -r^2p^2dv \, ,
\eea
and find that the line element (\ref{mPDf}) possesses the same form recently used in
\cite{LMP}.

The spacetime metric (\ref{EMCSKADS}) is of Petrov type $22$, like the three-equal-charge
Cveti\v{c}-Youm black hole solution and its super-symmetric counterpart --- the BMPV black
hole solution. It possesses a pair of real principal null vectors $\{\mathbf{l}, \mathbf{n}\}$,
a pair of complex principal null vectors $\{\mathbf{m}, \bar{\mathbf{m}}\}$, and one real,
spatial-like unit vector $\mathbf{k}$. They can be constructed to be of Kinnersley-type
as follows:
\begin{subequations}\begin{align}
\mathbf{l}^{\mu}\p_{\mu} &= \frac{1}{r^2\Delta_r}\Big[(r^2+a^2)(r^2+b^2)\p_t
 +(r^2+b^2)a\chi_a\p_{\phi} +(r^2+a^2)b\chi_b\p_{\psi} \nn \\
 &\quad +Q\big(ab\p_t +b\chi_a\p_{\phi} +a\chi_b\p_{\psi}\big)\Big] +\p_r \, , \\
\mathbf{n}^{\mu}\p_{\mu} &= \frac{1}{2r^2\Sigma}\Big[(r^2+a^2)(r^2+b^2)\p_t
 +(r^2+b^2)a\chi_a\p_{\phi} +(r^2+a^2)b\chi_b\p_{\psi} \nn \\
 &\quad +Q\big(ab\p_t +b\chi_a\p_{\phi} +a\chi_b\p_{\psi}\big)\Big]
  -\frac{\Delta_r}{2\Sigma}\p_r \, , \\
\mathbf{m}^{\mu}\p_{\mu} &= \frac{\sqrt{\Delta_p/2}}{r+ip}\bigg[\p_p
 +i\frac{(p^2-a^2)(p^2-b^2)}{p^2\Delta_p}\Big(\p_t
 -\frac{a\chi_a}{p^2-a^2}\p_{\phi} -\frac{b\chi_b}{p^2-b^2}\p_{\psi}\Big)\bigg] \, , \\
\bar{\mathbf{m}}^{\mu}\p_{\mu} &= \frac{\sqrt{\Delta_p/2}}{r-ip}\bigg[\p_p
 -i\frac{(p^2-a^2)(p^2-b^2)}{p^2\Delta_p}\Big(\p_t
 -\frac{a\chi_a}{p^2-a^2}\p_{\phi} -\frac{b\chi_b}{p^2-b^2}\p_{\psi}\Big)\bigg] \, , \\
\mathbf{k}^{\mu}\p_{\mu} &= \frac{1}{rp}\big(ab\p_t
 +b\chi_a\p_{\phi} +a\chi_b\p_{\psi}\big) \, .
\end{align}\end{subequations}
These vectors satisfy the following orthogonal relations
\be
\mathbf{l}^{\mu}\mathbf{n}_{\mu} = -1 \, ,  \qquad\quad
\mathbf{m}^{\mu}\bar{\mathbf{m}}_{\mu} = 1 \, ,
\qquad\quad \mathbf{k}^{\mu}\mathbf{k}_{\mu} = 1 \, ,
\label{Ognr}
\ee
and all others are zero. In terms of these vectors, the line element for the EMCS-Kerr-AdS
black hole (\ref{EMCSKADS}) can be put into a seminull pentad formalism ($2\bar{2}1$ formalism)
as follows:
\be
ds^2 = -\mathbf{l}\otimes \mathbf{n} -\mathbf{n}\otimes \mathbf{l} +\mathbf{m}\otimes
 \bar{\mathbf{m}} +\bar{\mathbf{m}}\otimes \mathbf{m} +\mathbf{k}\otimes \mathbf{k} \, .
\ee

\section{Hidden symmetries of Chong-Cveti\v{c}-L\"{u}-Pope
black hole and tower of generalized (conformal) Killing-Yano
tensors} \label{TGCKY}

In this section, we summarize our results about the complete hidden symmetry properties
of general five-dimensional Chong-Cveti\v{c}-L\"{u}-Pope black holes. In particular, we
propose how to generalize the concepts of Killing-Yano and conformal Killing-Yano tensors
so that they can be subject to the five-dimensional Einstein-Maxwell-Chern-Simons theory.
To proceed, we first give a brief review on the recent work about the construction of the
St\"{a}ckel-Killing tensor from the (conformal) Killing-Yano tensor.

Carter \cite{BC1} found that the geodesic Hamilton-Jacobi equation and Klein-Gordon scalar
field equation are separable by variables in the four-dimensional Kerr metric, and there
exists an additional quadratic integral of motion, called as the Carter's fourth constant.
This constant is associated with a second-order symmetric St\"{a}ckel-Killing tensor
$K_{\mu\nu} = K_{\nu\mu}$, which obeys the Killing equation
\be
K_{\mu\nu;\rho} +K_{\nu\rho;\mu} +K_{\rho\mu;\nu} = 0 \, .
\label{Kte}
\ee

Penrose and Floyd \cite{PF} further discovered that this St\"{a}ckel-Killing tensor can be
written in the form $K_{\mu\nu} = f_{\mu\rho} f^{~\rho}_{\nu}$, where the skew-symmetric
tensor $f_{\mu\nu} = -f_{\nu\mu}$ is the Killing-Yano tensor obeying the equation $f_{\mu\nu;
\rho} +f_{\mu\rho;\nu} = 0$. Using this object, Carter and McLenaghan \cite{CM} constructed
a first-order symmetry operator that commutes with the massive Dirac operator. In the case
of a $D = 4$ Kerr black hole, the Killing-Yano tensor $f$ is of rank-2, its Hodge dual
$k = -{^*}f$ is a rank-2, antisymmetric, conformal Killing-Yano tensor obeying the equation
\be
k_{\alpha\beta;\gamma} +k_{\alpha\gamma;\beta} = g_{\alpha\beta}\xi_{\gamma}
 +g_{\gamma\alpha}\xi_{\beta} -2g_{\beta\gamma}\xi_{\alpha} \, ,
\label{CKYe}
\ee
where the Killing vector is defined by
\be
\xi_{\alpha} = \frac{1}{D -1}k^{\mu}_{~\alpha;\mu} \, .
\ee
The above equation is equivalent to the Penrose's equation \cite{PenE}
\be
\mathcal{P}_{\alpha\beta\gamma} \equiv k_{\alpha\beta;\gamma}
 +g_{\beta\gamma}\xi_{\alpha} -g_{\gamma\alpha}\xi_{\beta} = 0 \, .
\ee

A conformal Killing-Yano tensor $k$ is dual to the Killing-Yano tensor if and only if
it is closed $dk = 0$. This fact implies that there at least locally exists a potential
one-form $\hat{b}$ so that $k = d\hat{b}$. Carter \cite{BC2} first found this potential
to generate the Killing-Yano tensor for the Kerr-Newman black hole.

Recently, these results have been extended to higher-dimensional rotating, uncharged
black hole solutions. In the special case of $D = 5$ dimensions, it was demonstrated
\cite{FKKF} that the rank-2 St\"{a}ckel-Killing tensor can be constructed from its ``square
root'', a rank-3, totally antisymmetric Killing-Yano tensor. According to Carter's recipe
\cite{BC2}, Frolov \textit{et al}. \cite{FKKF} started from a potential one-form to generate
a rank-2 conformal Killing-Yano tensor, whose Hodge dual $f = {^*}k$ is the expected rank-3
Killing-Yano tensor. The conformal Killing-Yano tensor $k = d\hat{b}$, the Killing-Yano
tensor $f = {^*}k$, and the Killing vector $h = {^*}(k\wedge k) = 2rp~e^5$ constitute a
tower of Killing-Yano tensors, and they are responsible for the whole hidden symmetries
of $D = 5$ Myers-Perry and Kerr-AdS black holes.

Now we focus on the general case of rotating, charged black holes in five dimensions. It
is clear that all of the five-dimensional Myers-Perry black hole, Kerr-AdS black hole,
three-equal-charge Cveti\v{c}-Youm black hole, and Chong-Cveti\v{c}-L\"{u}-Pope black
hole (\ref{EMCSKADS}) possess $R\times U(1)^2$ isometry group generated by three Killing
vectors ($\p_t$, $\p_{\phi}$, and $\p_{\psi}$). Besides, the separability properties of
the geodesic equation, the Hamilton-Jacobi equation, and the Klein-Gordon equation in
these black hole backgrounds imply that they are closely related to the existence of a
rank-2 symmetric St\"{a}ckel-Killing tensor admitted by all these spacetime metrics. In
the local Lorentzian coframe given in Eq. (\ref{pentad}), we \cite{WuMP,WuKAdS,WuCY} find
that the symmetric tensor $K_{\mu\nu} = K_{\nu\mu}$ in any of these spacetimes has a simple,
diagonal form
\be
K_{AB} = \mbox{diag} (-p^2, p^2, -r^2, -r^2, p^2-r^2) \, ,
\label{5dKT}
\ee
which is equivalent to those previously given in \cite{DKL,ANA1}, up to an additive constant.

On the other hand, it has been \cite{WuMP,WuKAdS} shown that the separability of the Dirac's
equation in the Myers-Perry and Kerr-AdS spacetime backgrounds is intimately associated with
the existence of a rank-3 antisymmetric Killing-Yano tensor admitted by these uncharged
metrics. What is more, it has been revealed that the symmetric St\"{a}ckel-Killing tensor
given by Eq. (\ref{5dKT}) can be written as the square of this rank-3 Killing-Yano tensor
\be
K_{\mu\nu} = -\frac{1}{2}f_{\mu\alpha\beta}f_{\nu}^{~\alpha\beta} \, ,
\label{sqroot}
\ee
where the rank-3 Killing-Yano tensor is given by
\be
f = \big(-p~e^0\wedge e^1 +r~e^2\wedge e^3\big)\wedge e^5 = {^*}k \, ,
\label{3KYt}
\ee
and satisfy the following Killing-Yano equation
\be
f_{\alpha\beta\mu;\nu} +f_{\alpha\beta\nu;\mu} = 0 \, .
\ee

The Hodge dual of the three-form $f$ is a two-form $k = -{^*}f$, which is a conformal
Killing-Yano tensor obeying Eq. (\ref{CKYe}). Adopting the following definitions:
\be
k_{\mu\nu} = -({^*}f)_{\mu\nu} = -\frac{1}{6}\sqrt{-g}
 \epsilon_{\mu\nu\alpha\beta\gamma}f^{\alpha\beta\gamma} \, , \qquad
f_{\alpha\beta\gamma} = ({^*}k)_{\alpha\beta\gamma} = \frac{1}{2}
 \sqrt{-g}\epsilon_{\alpha\beta\gamma\mu\nu}k^{\mu\nu} \, ,
\ee
and the convention $\epsilon^{01235} = 1 = -\epsilon_{01235}$ for the totally anti-symmetric
tensor density $\epsilon_{ABCDE}$, we find that this two-form is
\be
k = r~e^0\wedge e^1 +p~e^2\wedge e^3 = d\hat{b} \, ,
\label{cKYt}
\ee
which can be generated from a potential one-form
\be
2\hat{b} = \big(p^2-r^2\big)dt +\frac{(r^2+a^2)(p^2-a^2)a}{(b^2-a^2)\chi_a}d\phi
 +\frac{(r^2+b^2)(p^2-b^2)b}{(a^2-b^2)\chi_b}d\psi \, .
\label{KYp}
\ee
Clearly, the two-form $k = -{^*}f$ is closed, $dk = d^2\hat{b} = 0$, which indicates that
$f_{~~~~;\rho}^{\mu\nu\rho} = 0$.

If the conformal Killing-Yano tensor $k_{\mu\nu} = -k_{\nu\mu}$ is closed $dk = 0$, namely
\be
k_{\alpha\beta;\gamma} +k_{\beta\gamma;\alpha} +k_{\gamma\alpha;\beta} = 0 \, ,
\ee
then the Penrose potential possesses the following properties:
\begin{subequations}\begin{align}
& \mathcal{P}_{\alpha\beta\gamma} +\mathcal{P}_{\beta\alpha\gamma} = 0 \, , \\
& \mathcal{P}_{\alpha\beta\gamma} +\mathcal{P}_{\beta\gamma\alpha}
 +\mathcal{P}_{\gamma\alpha\beta} = 0 \, , \label{Ppcyc} \\
& \mathcal{P}_{\alpha\beta}^{~~~\beta} = 0 = \mathcal{P}_{\beta\alpha}^{~~~\beta} \, .
\end{align}\end{subequations}

We now demonstrate how the concepts of Killing-Yano and conformal Killing-Yano tensors can
be generalized to the charged case. Generally speaking, it is very complicated to find the
concrete expressions for the Killing objects via solving the equations that they should be
satisfied for the spacetime under consideration. Conversely, once given the analytical
expressions for the Killing objects from the beginning, it is relatively easy and simple to
check the equations that they obey. Therefore, we shall follow the latter routine. In other
words, we first assume that the expected Killing-Yano tensors have the same form as the one
in the uncharged case since they should recover it, then we examine their corresponding
properties and find out the new equation that they should fulfill.

Just like in the case of the three-equal-charge Cveti\v{c}-Youm black hole, the charged
Hamilton-Jacobi equation (essentially, the lowest order WKB approximation of Klein-Gordon
equation) and the complex Klein-Gordon equation for a scalar field with rest mass $\mu_0$
and electric charge $q$
\bea
&&\quad g^{\mu\nu}(\p_{\mu}\mathcal{S} +q\mathcal{A}_{\mu})(\p_{\nu}\mathcal{S}
 +q\mathcal{A}_{\nu}) +\mu_0^2 = 0 \, , \\
&& \big(\Box_c -\mu_0^2\big)\Phi = (\nabla_{\mu}+iq\mathcal{A}_{\mu})
 \big[g^{\mu\nu}(\nabla_{\nu} +iq\mathcal{A}_{\nu})\Phi\big] -\mu_0^2\Phi = 0 \, ,
\eea
are separable for variables in the Chong-Cveti\v{c}-L\"{u}-Pope black hole background geometry.
The separability of these equations indicates that the five-dimensional EMCS-Kerr-AdS metric
admits a rank-2 St\"{a}ckel-Killing tensor exactly given by Eq. (\ref{5dKT}). The separation
constant acts as the eigenvalue of the dual operator
\be
\mathbb{K}_c  = (\nabla_{\mu}+iq\mathcal{A}_{\mu})
 \big[K^{\mu\nu}(\nabla_{\nu} +iq\mathcal{A}_{\nu})\big] \, ,
\ee
which commutes with the complex scalar Laplacian operator $\Box_c$.

On the other hand, the separability of a spin-$1/2$ field equation \cite{WuCY}
\be
\big(\widetilde{\mathbb{H}}_D +\mu_e\big)\Psi = \Big[\gamma^{\mu}(\nabla_{\mu}
 +iq\mathcal{A}_{\mu}) +\frac{i}{4\sqrt{3}}\gamma^{\mu}\gamma^{\nu}F_{\mu\nu}
 +\mu_e\Big]\Psi = 0 \, ,
\ee
in the three-equal-charge Cveti\v{c}-Youm black hole and the Chong-Cveti\v{c}-L\"{u}-Pope
spacetime backgrounds is also closely associated with the existence of a rank-3 antisymmetric
tensor admitted by these spacetimes. The separation constant introduced in the modified Dirac
equation behaves as the eigenvalue of a first-order differential operator \cite{WuCY}
\be
\mathbb{H}_f = -\frac{1}{2}\gamma^{\mu}\gamma^{\nu}f^{~~\rho}_{\mu\nu}
 \big(\nabla_{\rho} +iq\mathcal{A}_{\rho}\big) +\frac{1}{16}\gamma^{\mu}
  \gamma^{\nu}\gamma^{\rho}\gamma^{\sigma}f_{\mu\nu\rho;\sigma} \, ,
\label{dualDiracop}
\ee
that commutes with the modified Dirac operator $\widetilde{\mathbb{H}}_D$. Here the minimal
electro-magnetic coupling interaction has been taken into consideration.

Now that in the local Lorentzian coframe pentad (\ref{pentad}), the St\"{a}ckel-Killing
tensor has a simple, diagonal form given by Eq. (\ref{5dKT}), it is reasonable to assume
that the expected anti-symmetric tensor of rank-3 is still given by Eq. (\ref{3KYt})
which can naturally reduce to that in the uncharged case \cite{WuMP,WuKAdS}. What is more,
this anti-symmetric tensor still can be understood as the ``square root'' of the rank-2
St\"{a}ckel-Killing tensor via Eq. (\ref{sqroot}). It is also easy to find that the Hodge
dual of this rank-3 tensor is still given by Eq. (\ref{cKYt}) and can be generated from
a potential one-form (\ref{KYp}), similar to the uncharged Myers-Perry and Kerr-AdS black
hole cases.

At this stage, we are in a position to check whether these anti-symmetric tensors still
satisfy the usual (conformal) Killing-Yano equation. In our previous work \cite{WuCY},
we have proposed to generalize the concepts of the (conformal) Killing-Yano tensors so
that they can be subject to the five-dimensional minimal gauged supergravity theory. To
this end, we propose that a generalized conformal Killing-Yano tensor should satisfy
the following equation
\be
\mathcal{P}_{\alpha\beta\gamma}
 = \frac{1}{\sqrt{3}}\widetilde{F}_{\alpha\beta}^{~~\lambda}k_{\gamma\lambda}
 = \frac{1}{\sqrt{3}}f_{\alpha\beta}^{~~\lambda}F_{\gamma\lambda} \, .
\label{GcKYe}
\ee
Equivalently, it can be rewritten as
\be
k_{\alpha\beta;\gamma} +k_{\alpha\gamma;\beta} = g_{\alpha\beta}\xi_{\gamma}
 +g_{\gamma\alpha}\xi_{\beta} -2g_{\beta\gamma}\xi_{\alpha}
 +\frac{1}{\sqrt{3}}\big(\widetilde{F}_{\alpha\beta}^{~~\lambda}k_{\gamma\lambda}
 +\widetilde{F}_{\alpha\gamma}^{~~\lambda}k_{\beta\lambda}\big) \, ,
\ee
where $4\xi^{\alpha} = k^{\mu\alpha}_{~~;\mu}$, and the dual Maxwell three-form is defined by
\be
\widetilde{F}_{\alpha\beta\gamma} = ({^*}F)_{\alpha\beta\gamma} = \frac{1}{2}
 \sqrt{-g}\epsilon_{\alpha\beta\gamma\mu\nu}F^{\mu\nu} \, .
\ee
The cyclic property of the Penrose potential leads to the following important identities
\be
k \wedge F = 0 = f\wedge \widetilde{F} \, .
\ee

Since this rank-3 antisymmetric tensor $f = {^*}k$ is the Hodge dual of the two-form $k =
d\hat{b}$, we can take the Hodge dual of the generalized Penrose equation (\ref{GcKYe})
and obtain
\bea
f_{\alpha\beta\mu;\nu} &=& \frac{1}{2}\sqrt{-g}\epsilon_{\alpha\beta\mu\rho\sigma}
 k^{\rho\sigma}_{~~;\nu} \nn \\
 &=& \frac{1}{2}\sqrt{-g}\epsilon_{\alpha\beta\mu\rho\sigma}
 \big(\mathcal{P}^{\rho\sigma}_{~~\nu} +\delta_{\nu}^{\rho}\xi^{\sigma}
 -\delta_{\nu}^{\sigma}\xi^{\rho}\big) \nn \\
&=& \mathcal{W}_{\alpha\beta\mu\nu}
 +\sqrt{-g}\epsilon_{\alpha\beta\mu\nu\sigma}\xi^{\sigma} \, ,
\label{gcPe}
\eea
where we have denoted
\be
\mathcal{W}_{\alpha\beta\mu\nu} \equiv
  \frac{1}{2}\sqrt{-g}\epsilon_{\alpha\beta\mu\rho\sigma}\mathcal{P}^{\rho\sigma}_{~~\nu}
 = \frac{1}{2\sqrt{3}}\sqrt{-g}\epsilon_{\alpha\beta\mu\rho\sigma}
 f^{\rho\sigma\lambda}F_{\nu\lambda} \, .
\ee
Symmetrization of Eq. (\ref{gcPe}) with respect to the last two indices ($\mu, \nu$) gives
a generalized Killing-Yano equation
\be
f_{\alpha\beta\mu;\nu} +f_{\alpha\beta\nu;\mu}
= \mathcal{W}_{\alpha\beta\mu\nu} +\mathcal{W}_{\alpha\beta\nu\mu} \, ,
\label{gKYe}
\ee
which had already been proposed in our previous work \cite{WuCY}.

Finally, after a lengthy computations, we can work out the commutator
\bea
[\mathbb{H}_f, \widetilde{\mathbb{H}}_D]
&=& \Big[\frac{1}{8}\gamma^{\alpha}\gamma^{\beta}\gamma^{\nu}\big(\nabla_{\nu}
 f^{~~\mu}_{\alpha\beta} +\nabla^{\mu}f_{\alpha\beta\nu}\big)
 +\frac{i}{2\sqrt{3}}\big(\gamma^{\nu}\gamma^{\beta}
 -\gamma^{\beta}\gamma^{\nu}\big)F_{\nu}^{~\alpha}f^{~~\mu}_{\alpha\beta}\Big]
 \big(\nabla_{\mu} +iq\mathcal{A}_{\mu}\big) \nn \\
&& -\frac{1}{16}\gamma^{\mu}\gamma^{\alpha}\gamma^{\beta}\gamma^{\rho}
 \gamma^{\sigma}\big(\nabla_{\mu}\nabla_{\sigma}f_{\alpha\beta\rho}
 +2f^{~~\nu}_{\mu\beta}R_{\rho\sigma\alpha\nu}\big)
  -\frac{iq}{2}\gamma^{\mu}\gamma^{\alpha}\gamma^{\beta}
  f^{~~\nu}_{\mu\beta}F_{\alpha\nu} \nn \\
&& +\frac{i}{16\sqrt{3}}\gamma^{\alpha}\gamma^{\beta}\gamma^{\rho}\gamma^{\sigma}
 \big(3F_{~\alpha}^{\mu}\nabla_{\sigma}f_{\mu\beta\rho} +F_{~\sigma}^{\mu}\nabla_{\mu}
 f_{\alpha\beta\rho} -2f^{~~\mu}_{\alpha\beta}\nabla_{\mu}F_{\rho\sigma}\big) \nn \\
&& -\frac{\sqrt{3}i}{8}\gamma^{\rho}\gamma^{\sigma}
 F^{\mu\nu}\nabla_{\sigma}f_{\mu\nu\rho} \, .
\eea
To derive the last expression for this commutator, we have made use of the anti-commutativity
of Dirac gamma matrices and the following relations
\be
\nabla_{\mu}\gamma^{\nu} = 0 \, , \qquad [\nabla_{\mu}, \nabla_{\nu}] =
 \frac{1}{4}R_{\mu\nu\rho\sigma}\gamma^{\rho}\gamma^{\sigma} \, , \qquad
 f^{\rho}_{~\mu\nu;\rho} = 0 \, .
\ee

Then we can see that the commutation relation $[\mathbb{H}_f, \widetilde{\mathbb{H}}_D] = 0$
just yields the generalized Killing-Yano equation (\ref{gKYe}) and the integrability condition
for this generalized Killing-Yano tensor of rank-3 .

\section{Separability of a massive Klein-Gordon complex scalar field
equation and rank-2 St\"{a}ckel-Killing tensor} \label{KGSK}

In this section, the massive Klein-Gordon complex scalar field equation with a minimal
gauge coupling term is separated by variables into purely radial and purely angular parts
in the five-dimensional EMCS-Kerr-AdS black hole geometry. From the separated solutions,
we can construct a second-order operator that commutes with the complex scalar Laplacian
operator. We then show that this second-order symmetry operator can be compactly expressed
in terms of a rank-2 symmetric St\"{a}ckel-Killing tensor which has a simple, diagonal form
in the chosen local Lorentzian coframe one-forms (\ref{pentad}).

To begin with, let us consider a massive Klein-Gordon complex scalar field equation with
a minimal electro-magnetic interaction
\be
\big(\Box_c -\mu_0^2\big)\Phi = \frac{1}{\sqrt{-g}}(\p_{\mu}+iq\mathcal{A}_{\mu})
 \big[\sqrt{-g} g^{\mu\nu}(\p_{\nu}+iq\mathcal{A}_{\nu})\Phi\big] -\mu_0^2\Phi = 0 \, .
\ee

The metric determinant for the spacetime (\ref{mPDf}) is
\be
\sqrt{-g} = \frac{rp\Sigma}{(a^2-b^2)\chi_a\chi_b} \, ,
\ee
and the contra-invariant components for the metric tensor can be read accordingly from
\bea
g^{\mu\nu}\p_{\mu}\p_{\nu} &=& \eta^{AB}\p_A\otimes\p_B \nn \\
&=& -\frac{1}{r^4\Delta_r\Sigma}\Big[(r^2+a^2)(r^2+b^2)\p_t
 +(r^2+b^2)a\chi_a\p_{\phi} +(r^2+a^2)b\chi_b\p_{\psi} \nn \\
&& +Q\big(ab\p_t +b\chi_a\p_{\phi} +a\chi_b\p_{\psi}\big)\Big]^2
 +\frac{\Delta_r}{\Sigma}\p_r^2 +\frac{\Delta_p}{\Sigma}\p_p^2 \nn \\
&& +\frac{(p^2-a^2)^2(p^2-b^2)^2}{p^4\Delta_p\Sigma}
\Big(\p_t -\frac{a\chi_a}{p^2-a^2}\p_{\phi} -\frac{b\chi_b}{p^2-b^2}\p_{\psi}\Big)^2 \nn \\
&&\quad +\frac{1}{r^2p^2}\big(ab\p_t +b\chi_a\p_{\phi} +a\chi_b\p_{\psi}\big)^2 \, .
\eea

In the background spacetime (\ref{mPDf}), the massive complex scalar field equation reads
\bea
&&\bigg\{ -\frac{1}{r^4\Delta_r\Sigma}\Big[(r^2+a^2)(r^2+b^2)\p_t
 +(r^2+b^2)a\chi_a\p_{\phi} +(r^2+a^2)b\chi_b\p_{\psi} \nn \\
&&\qquad +Q\big(ab\p_t +b\chi_a\p_{\phi} +a\chi_b\p_{\psi}\big)
 +\frac{\sqrt{3}}{2}iqQr^2\Big]^2 +\frac{1}{r\Sigma}\p_r\big(r\Delta_r\p_r\big) \nn \\
&&\qquad +\frac{1}{p\Sigma}\p_p\big(p\Delta_p\p_p\big)
 +\frac{(p^2-a^2)^2(p^2-b^2)^2}{p^4\Delta_p\Sigma}\Big(\p_t
 -\frac{a\chi_a}{p^2-a^2}\p_{\phi} -\frac{b\chi_b}{p^2-b^2}\p_{\psi}\Big)^2 \nn \\
&&\qquad\quad +\frac{1}{r^2p^2}\big(ab\p_t +b\chi_a\p_{\phi} +a\chi_b\p_{\psi}\big)^2
 -\mu_0^2\bigg \}\Phi = 0 \, .
\eea
With the ansatz of separation of variables $\Phi = R(r)S(p)e^{i(m\phi +k\psi -\omega t)}$,
it apparently can be separated into a radial part and an angular part,
\bea
&& \frac{1}{r}\p_r\big(r\Delta_r\p_rR\big) +\Big\{\frac{1}{r^4\Delta_r}
 \Big[(r^2+a^2)(r^2+b^2)\omega -(r^2+b^2)ma\chi_a \nn \\
&&\qquad\qquad -(r^2+a^2)kb\chi_b +Q\big(ab\omega -mb\chi_a -ka\chi_b\big)
 -\frac{\sqrt{3}}{2}qQr^2\Big]^2 \nn \\
&&\qquad\qquad\quad -\frac{1}{r^2}\big(ab\omega -mb\chi_a -ka\chi_b\big)^2
 -\mu_0^2r^2 -\lambda_0^2 \Big\}R(r) = 0 \, , \label{srs} \\
&& \frac{1}{p}\p_p\big(p\Delta_p\p_pS\big)
 -\Big\{\frac{(p^2-a^2)^2(p^2-b^2)^2}{p^4\Delta_p}\Big(\omega
 +\frac{ma\chi_a}{p^2-a^2} +\frac{kb\chi_b}{p^2-b^2}\Big)^2 \nn \\
&&\qquad\qquad\qquad\quad  +\frac{1}{p^2}\big(ab\omega -mb\chi_a -ka\chi_b\big)^2
 +\mu_0^2p^2 -\lambda_0^2\Big\}S(p) = 0 \, . \label{sra}
\eea
Both of them can be transformed into the general form of Heun equation \cite{Heun}.

Now from the separated parts (\ref{srs}) and (\ref{sra}), one can construct a new dual
equation as follows:
\bea
&&\bigg\{ -\frac{p^2}{r^4\Delta_r\Sigma}\Big[(r^2+a^2)(r^2+b^2)\p_t
 +(r^2+b^2)a\chi_a\p_{\phi} +(r^2+a^2)b\chi_b\p_{\psi}  +\frac{\sqrt{3}}{2}iqQr^2 \nn \\
&&\quad +Q\big(ab\p_t +b\chi_a\p_{\phi} +a\chi_b\p_{\psi}\big)\Big]^2
 +\frac{p^2}{r\Sigma}\p_r\big(r\Delta_r\p_r\big)
 -\frac{r^2}{p\Sigma}\p_p\big(p\Delta_p\p_p\big) \nn \\
&&\qquad -r^2\frac{(p^2-a^2)^2(p^2-b^2)^2}{p^4\Delta_p\Sigma}\Big(\p_t
 -\frac{a\chi_a}{p^2-a^2}\p_{\phi} -\frac{b\chi_b}{p^2-b^2}\p_{\psi}\Big)^2 \nn \\
&&\qquad\quad +\frac{p^2-r^2}{r^2p^2}\big(ab\p_t +b\chi_a\p_{\phi}
 +a\chi_b\p_{\psi}\big)^2 -\lambda_0^2\bigg \}\Phi = 0 \, ,
\eea
from which by setting $q = 0$ we can extract a second-order symmetric tensor --- the
so-called St\"{a}ckel-Killing tensor
\bea
K^{\mu\nu}\p_{\mu}\p_{\nu} &=& -p^2\frac{1}{r^4\Delta_r\Sigma} \Big[(r^2+a^2)(r^2+b^2)\p_t
 +(r^2+b^2)a\chi_a\p_{\phi} +(r^2+a^2)b\chi_b\p_{\psi} \nn \\
&& +Q\big(ab\p_t +b\chi_a\p_{\phi} +a\chi_b\p_{\psi}\big)\Big]^2
 +p^2\frac{\Delta_r}{\Sigma}\p_r^2 -r^2\frac{\Delta_p}{\Sigma}\p_p^2 \nn \\
&& -r^2\frac{(p^2-a^2)^2(p^2-b^2)^2}{p^4\Delta_p\Sigma}\Big(\p_t
 -\frac{a\chi_a}{p^2-a^2}\p_{\phi} -\frac{b\chi_b}{p^2-b^2}\p_{\psi}\Big)^2 \nn \\
&&\quad +\frac{p^2-r^2}{r^2p^2}\big(ab\p_t +b\chi_a\p_{\phi} +a\chi_b\p_{\psi}\big)^2 \, .
\eea
This rank-2 tensor has a concise expression given by Eq. (\ref{5dKT}) in terms of the
local Lorentzian orthonormal pentad (\ref{pentad}).

Using this St\"{a}ckel-Killing tensor, the above dual equation can be put into an operator
form
\be
\big(\mathbb{K}_c -\lambda_0^2\big)\Phi = \frac{1}{\sqrt{-g}}(\p_{\mu}+iq\mathcal{A}_{\mu})
 \big[\sqrt{-g} K^{\mu\nu}(\p_{\nu}+iq\mathcal{A}_{\nu})\Phi\big] -\lambda_0^2\Phi = 0 \, .
\ee
Clearly, the symmetry operator $\mathbb{K}_c$ is expressed in terms of the St\"{a}ckel-Killing
tensor and commutes with the complex scalar Laplacian operator $\Box_c$. The introduced
separation constant $\lambda_0^2$ acts as the eigenvalue of this operator. Expanding the
commutator $[\mathbb{K}_c, \Box_c] = 0$ yields the Killing equation (\ref{Kte}) and the
integrability condition for the St\"{a}ckel-Killing tensor.

\section{Separability of a complex massive spinor field equation in the
$D = 5$ Chong-Cveti\v{c}-L\"{u}-Pope black hole background} \label{SmDE}

In our previous work \cite{WuCY}, the modified Dirac equation for spin-$1/2$ fermions in
the three-equal-charge Cveti\v{c}-Youm black hole geometry has been decoupled into purely
radial and purely angular parts by using a local orthonormal f\"{u}nfbein (pentad) formalism.
In this section, we shall extend that work to the case of a nonzero cosmological constant
by showing that our modified Dirac equation is separable by variables in the $D = 5$
Chong-Cveti\v{c}-L\"{u}-Pope black hole background geometry.

\subsection{F\"{u}nfbein formalism of spinor field equation}

In Ref. \cite{WuCY}, it has been shown that in order to separate the field equation for
spin-$1/2$ fermions in a fixed background spacetime subject to the five-dimensional EMCS
supergravity theory, an extra counter-term should be supplemented to the usual Dirac
equation. The same thing holds true in minimal $D = 5$ gauged supergravity theory.
Incorporating the minimal electro-magnetic coupling interaction, the action of spin-$1/2$
spinor particles is therefore given by
\be
S_f = \frac{i}{2}\int d^5x\sqrt{-g}~\overline{\Psi}\Big[\gamma^{\mu}\big(\nabla_{\mu}
 +iq\mathcal{A}_{\mu}\big) +\frac{i}{4\sqrt{3}}\gamma^{\mu}\gamma^{\nu}F_{\mu\nu}
 +\mu_e\Big]\Psi \, ,
\ee
where $\Psi$ is a complex four-component Dirac spinor, $\mu_e$ is the rest mass, and $q$
is the charge of the electron.

Variation of the above action with respect to the spinor field yields the equation of motion
\be
\big(\widetilde{\mathbb{H}}_D +\mu_e\big)\Psi =
 \Big[\gamma^Ae_A^{~\mu}(\p_{\mu} +\Gamma_{\mu} +iq\mathcal{A}_{\mu})
 +\frac{i}{4\sqrt{3}}\gamma^A\gamma^BF_{AB} +\mu_e\Big]\Psi = 0 \, ,
\label{mDE}
\ee
where $e_A^{~\mu}$ is the f\"{u}nfbein (pentad), its inverse $e_{~\mu}^A$ is defined by
$g_{\mu\nu} = \eta_{AB}e_{~\mu}^Ae_{~\nu}^B$, $\Gamma_{\mu}$ is the spinor connection,
and $\gamma^A$'s are the five-dimensional gamma matrices obeying the anticommutation
relations (Clifford algebra)
\be
\big\{\gamma^A, \gamma^B\big\} \equiv \gamma^A\gamma^B +\gamma^B\gamma^A = 2\eta^{AB} \, .
\label{Clifford}
\ee
To our aim, we choose the following explicit representations for the gamma matrices
\cite{WuMP}
\bea
&& \gamma^0 = i\sigma^1\otimes I_2 \, , \qquad\quad \gamma^1 = -\sigma^2\otimes \sigma^3 \, ,
\qquad\quad \gamma^2 = -\sigma^2\otimes \sigma^1 \, , \nn \\
&& \gamma^3 = -\sigma^2\otimes \sigma^2 \, , \qquad\qquad\quad
\gamma^5 = \sigma^3\otimes I_2 = -i\gamma^0\gamma^1\gamma^2\gamma^3 \, ,
\label{GMr}
\eea
where $\sigma^i$'s are the Pauli matrices, and $I_2$ is a $2 \times 2$ identity matrix,
respectively.

In the f\"{u}nfbein formalism, the modified Dirac field equation (\ref{mDE}) can be
rewritten in the local Lorentzian frame as \cite{WuCY}
\be
\big(\widetilde{\mathbb{H}}_D +\mu_e\big)\Psi = \Big[\gamma^A(\p_A +\Gamma_A
 +iq\mathcal{A}_{A}) +\frac{i}{4\sqrt{3}}\gamma^A\gamma^BF_{AB} +\mu_e\Big]\Psi = 0 \, ,
\ee
where $\p_A = e_A^{~\mu}\p_{\mu}$ is the local partial differential operator and $\Gamma_A
= e_A^{~\mu}\Gamma_{\mu}$ is the component of the spinor connection projected in the local
Lorentzian frame. Note that the five-dimensional Clifford algebra has two different but
reducible representations which can differ by the multiplier of a $\gamma^5$ matrix. It
is usually assumed that fermion fields are in a reducible representation of the Clifford
algebra.

\subsection{Computation of covariant spinor differential operator}

In order to get the explicit expression of the modified Dirac's equation in the local
Lorentzian frame, one has to find firstly the local partial differential operator $\p_A
= e_A^{~\mu}\p_{\mu}$ and the spinor connection $\Gamma_A = e_A^{~\mu}\Gamma_{\mu}$
subject to the background metric (\ref{mPDf}). Once the pentad coframe one-forms $e^A
= e^A_{~\mu}dx^{\mu}$ have been concretely chosen, the local differential operator $\p_A
= e_A^{~\mu}\p_{\mu}$ can be determined via the orthogonal relations: $e_A^{~\mu}
e^B_{~\mu} = \delta_A^B$ and $e_A^{~\mu}e^A_{~\nu} = \delta^{\mu}_{\nu}$.

The orthonormal basis one-vectors $\p_A $ dual to the pentad $e^{A}$ given in Eq.
(\ref{pentad}) are given by
\bea
&& \p_0 = \frac{1}{r^2\sqrt{\Delta_r\Sigma}}\Big[(r^2+a^2)(r^2+b^2)\p_t
 +(r^2+b^2)a\chi_a\p_{\phi} \nn \\
&&\qquad +(r^2+a^2)b\chi_b\p_{\psi} +Q\big(ab\p_t +b\chi_a\p_{\phi}
 +a\chi_b\p_{\psi}\big)\Big] \, , \nn \\
&& \p_1 = \sqrt{\frac{\Delta_r}{\Sigma}}\p_r \, , \qquad\qquad
 \p_2 = \sqrt{\frac{\Delta_p}{\Sigma}}\p_p \, , \nn \\
&& \p_3 = \frac{(p^2-a^2)(p^2-b^2)}{p^2\sqrt{\Delta_p\Sigma}}\Big(\p_t
 -\frac{a\chi_a}{p^2-a^2}\p_{\phi} -\frac{b\chi_b}{p^2-b^2}\p_{\psi}\Big) \, , \nn \\
&& \p_5 = \frac{1}{rp}\big(ab\p_t +b\chi_a\p_{\phi} +a\chi_b\p_{\psi}\big) \, .
\eea
Therefore, the spinor partial differential operator is
\bea
&& \gamma^A\p_A = \gamma^0 \frac{1}{r^2\sqrt{\Delta_r\Sigma}}\Big[(r^2+a^2)(r^2+b^2)\p_t
 +(r^2+b^2)a\chi_a\p_{\phi} +(r^2+a^2)b\chi_b\p_{\psi} \nn \\
&&\qquad\qquad +Q\big(ab\p_t +b\chi_a\p_{\phi} +a\chi_b\p_{\psi}\big)\Big]
 +\gamma^1\sqrt{\frac{\Delta_r}{\Sigma}}\p_r
 +\gamma^2\sqrt{\frac{\Delta_p}{\Sigma}}\p_p \nn \\
&&\qquad\qquad +\gamma^3 \frac{(p^2-a^2)(p^2-b^2)}{p^2\sqrt{\Delta_p\Sigma}}\Big(\p_t
 -\frac{a\chi_a}{p^2-a^2}\p_{\phi} -\frac{b\chi_b}{p^2-b^2}\p_{\psi}\Big) \nn \\
&&\qquad\qquad +\gamma^5\frac{1}{rp}\big(ab\p_t +b\chi_a\p_{\phi} +a\chi_b\p_{\psi}\big) \, ,
\eea

The next step is to compute the component $\Gamma_A$ of the spinor connection. The procedure
of the detailed computations is outlined in the appendix. For our purpose, we need the final
expression
\bea
\gamma^A\Gamma_A &=&
 \gamma^1\sqrt{\frac{\Delta_r}{\Sigma}}\Big(\frac{\Delta_r^{\prime}}{4\Delta_r}
 +\frac{1}{2r} +\frac{r -ip\gamma^5}{2\Sigma}\Big) +\gamma^2\sqrt{\frac{\Delta_p}{\Sigma}}
 \Big(\frac{\Delta_p^{\prime}}{4\Delta_p} +\frac{1}{2p}
 +\frac{p +ir\gamma^5}{2\Sigma}\Big) \nn \\
&& +\Big(\frac{Q+ab}{2r^2\Sigma} +\frac{ab}{2p^2\Sigma}\Big)i\gamma^0\gamma^1
 \big(r +ip\gamma^5\big) -\frac{Q}{2\Sigma^2}\big(ir\gamma^0\gamma^1
  +p\gamma^0\gamma^1\gamma^5\big) \, ,
\eea
where a prime denotes the partial differential with respect to the coordinates $r$ and $p$.

As explained out in \cite{WuCY}, the appearance of last term in the expression of $\gamma^A
\Gamma_A$ spoils the separability of the usual Dirac equation. To cancel it, one should
supplement an additional term
\be
\frac{i}{4\sqrt{3}}\gamma^A\gamma^BF_{AB}
= \frac{iQ}{2\Sigma^2}\big(r\gamma^0\gamma^1 -p\gamma^2\gamma^3\big)
\equiv \frac{Q}{2\Sigma^2}\big(p\gamma^0\gamma^1 +r\gamma^2\gamma^3\big)\gamma^5
= \frac{1}{12\sqrt{3}}\gamma^A\gamma^B\gamma^C\widetilde{F}_{ABC} \, .
\ee
With the inclusion of this new counter-term, then the modified Dirac equation for a spin-$1/2$
spinor field in this general rotating, charged, EMCS-Kerr-AdS black hole spacetime can be
completely decoupled into purely radial and purely angular parts. Without this counter-term,
the usual Dirac equation is only separable \cite{DKL} in the special case when $a =\pm b$.
It should be emphasized that the work on the separation of Dirac equation in \cite{DKL} is
incomplete, since our counter-term still make a contribution to the Dirac equation in the
$a =\pm b$ case.

Combining the above expressions with the minimal gauge coupling factor
\be
iq\gamma^{\mu}\mathcal{A}_{\mu} = iq\gamma^A\mathcal{A}_{A}
 = \gamma^0\frac{\sqrt{3}iqQ}{2\sqrt{\Delta_r\Sigma}} \, ,
\ee
we find that the modified Dirac's covariant differential operator in the local Lorentzian
frame is
\bea
\widetilde{\mathbb{H}}_D
&=& \gamma^0\frac{1}{r^2\sqrt{\Delta_r\Sigma}}\Big[(r^2+a^2)(r^2+b^2)\p_t
 +(r^2+b^2)a\chi_a\p_{\phi} +(r^2+a^2)b\chi_b\p_{\psi} \nn \\
&& +Q\big(ab\p_t +b\chi_a\p_{\phi} +a\chi_b\p_{\psi}\big) +\frac{\sqrt{3}}{2}iqQr^2 \Big]
 +\gamma^1\sqrt{\frac{\Delta_r}{\Sigma}}\Big(\p_r +\frac{\Delta_r^{\prime}}{4\Delta_r}
 +\frac{1}{2r} +\frac{r -ip\gamma^5}{2\Sigma}\Big) \nn \\
&& +\gamma^2\sqrt{\frac{\Delta_p}{\Sigma}}\Big[\p_p +\frac{\Delta_p^{\prime}}{4\Delta_p}
 +\frac{1}{2p} +\frac{i\gamma^5}{2\Sigma}\big(r -ip\gamma^5\big)\Big]
 +\gamma^3\frac{(p^2-a^2)(p^2-b^2)}{p^2\sqrt{\Delta_p\Sigma}}\Big(\p_t
 -\frac{a\chi_a}{p^2-a^2}\p_{\phi} \nn \\
&& -\frac{b\chi_b}{p^2-b^2}\p_{\psi}\Big) +\gamma^5\frac{1}{rp}\big(ab\p_t
 +b\chi_a\p_{\phi} +a\chi_b\p_{\psi}\big) +\Big(\frac{Q+ab}{2r^2\Sigma}
 +\frac{ab}{2p^2\Sigma}\Big)i\gamma^0\gamma^1\big(r +ip\gamma^5\big) \, . \nn \\ &&
\eea

\subsection{Separation of variables of spinor field equation}

With the above preparation in hand, we are ready to decouple the modified Dirac equation
\be
\big(\widetilde{\mathbb{H}}_D +\mu_e\big)\Psi
 = \Big[\gamma^{\mu}(\p_{\mu} +\Gamma_{\mu} +iq\mathcal{A}_{\mu})
 +\frac{i}{4\sqrt{3}}\gamma^{\mu}\gamma^{\nu}F_{\mu\nu} +\mu_e\Big]\Psi = 0 \, .
\label{DELF}
\ee
Substituting the above spinor covariant differential operator into Eq. (\ref{DELF}) and
multiplying it a factor $(r -ip\gamma^5)\sqrt{r +ip\gamma^5} = \sqrt{\Sigma(r -ip\gamma^5)}$
by the left, then after some lengthy algebra manipulations we get the following expression
for the modified Dirac equation in the five-dimensional EMCS-Kerr-AdS metric
\bea
&& \bigg\{ \gamma^0\frac{1}{r^2\sqrt{\Delta_r}}\Big[(r^2+a^2)(r^2+b^2)\p_t
+(r^2+b^2)a\chi_a\p_{\phi} +(r^2+a^2)b\chi_b\p_{\psi} \nn \\
&&\qquad +Q\big(ab\p_t +b\chi_a\p_{\phi} +a\chi_b\p_{\psi}\big)
 +\frac{\sqrt{3}}{2}iqQr^2\Big] +\gamma^1\sqrt{\Delta_r}\Big(\p_r
 +\frac{\Delta_r^{\prime}}{4\Delta_r} +\frac{1}{2r}\Big) \nn \\
&&\quad +\gamma^2\sqrt{\Delta_p}\Big(\p_p +\frac{\Delta_p^{\prime}}{4\Delta_p}
 +\frac{1}{2p}\Big) +\gamma^3\frac{(p^2-a^2)(p^2-b^2)}{p^2\sqrt{\Delta_p}}\Big(\p_t
 -\frac{a\chi_a}{p^2-a^2}\p_{\phi} -\frac{b\chi_b}{p^2-b^2}\p_{\psi}\Big) \nn \\
&&\quad +\Big(\frac{\gamma^5}{p} -\frac{i}{r}\Big)\big(ab\p_t +b\chi_a\p_{\phi}
 +a\chi_b\p_{\psi}\big) +\Big(\frac{Q+ab}{2r^2} +\frac{ab}{2p^2}\Big)i\gamma^0\gamma^1 \nn \\
&&\qquad  +\mu_e\big(r -ip\gamma^5\big) \bigg\}\big(\sqrt{r +ip\gamma^5}\Psi\big) = 0 \, .
\label{presde}
\eea

Now we assume that the spin-$1/2$ fermion fields are in a reducible representation of the
Clifford algebra and can be taken as a complex four-component Dirac spinor. Adopting the
explicit representation (\ref{GMr}) for the gamma matrices and the following ansatz for
the separation of variables
\be
\sqrt{r +ip\gamma^5}\Psi = e^{i(m\phi+k\psi-\omega t)}\left(\ba{cl}
&\hspace*{-5pt} R_2(r)S_1(p) \\
&\hspace*{-5pt} R_1(r)S_2(p) \\
&\hspace*{-5pt} R_1(r)S_1(p) \\
&\hspace*{-5pt} R_2(r)S_2(p)
\ea\right) \, ,
\ee
we find that the modified Dirac equation in the five-dimensional EMCS-Kerr-AdS metric can
be decoupled into the purely radial parts and the purely angular parts
\bea
&& \sqrt{\Delta_r}\cD_r^-R_1 = \Big[\lambda +i\mu_er -\frac{Q+ab}{2r^2}
 -\frac{i}{r}\big(ab\omega -mb\chi_a -ka\chi_b\Big)\big]R_2 \, , \label{sdea} \\
&& \sqrt{\Delta_r}\cD_r^+R_2 = \Big[\lambda -i\mu_er -\frac{Q+ab}{2r^2}
+\frac{i}{r}\big(ab\omega -mb\chi_a -ka\chi_b\big)\Big]R_1 \, , \label{sdeb} \\
&& \sqrt{\Delta_p}\cL_p^+S_1 = \Big[\quad\lambda +\mu_ep +\frac{ab}{2p^2}
 +\frac{1}{p}\big(ab\omega -mb\chi_a -ka\chi_b\big)\Big]S_2 \, , \label{sdec} \\
&& \sqrt{\Delta_p}\cL_p^-S_2 = \Big[-\lambda +\mu_ep -\frac{ab}{2p^2}
 +\frac{1}{p}\big(ab\omega -mb\chi_a -ka\chi_b\big)\Big]S_1 \label{sded} \, ,
\eea
in which $\lambda$ is the separation constant, and we have introduced
\bea
&& \cD_r^{\pm} = \p_r +\frac{\Delta_r^{\prime}}{4\Delta_r} +\frac{1}{2r}
 \pm i\frac{1}{r^2\Delta_r} \Big[(r^2+a^2)(r^2+b^2)\omega -(r^2+b^2)ma\chi_a \nn \\
&&\qquad\qquad  -(r^2+a^2)kb\chi_b +Q\big(ab\omega -mb\chi_a -ka\chi_b\big)
 -\frac{\sqrt{3}}{2}qQr^2\Big] \, , \nn \\
&& \cL_p^{\pm} = \p_p +\frac{\Delta_p^{\prime}}{4\Delta_p} +\frac{1}{2p} \pm
  \frac{(p^2-a^2)(p^2-b^2)}{p^2\Delta_p}\Big(\omega +\frac{ma\chi_a}{p^2-a^2}
  +\frac{kb\chi_b}{p^2-b^2}\Big) \, . \nn
\eea

The separated radial and angular equations (\ref{sdea}-\ref{sded}) can be reduced into
a master equation containing only one component. The decoupled master equations are very
complicated and not given here. As for the exact solution to these equations, one hopes
that they can be transformed into the general form of Heun equation \cite{Heun}. Besides,
the angular part can be transformed into the radial part if one replaces $p$ by $ir$ in
the vacuum case when $M = Q = 0$.

\section{Separability of dual first-order differential operator
equation in terms of generalized Killing-Yano tensor} \label{dEgKY}

In the last section, we have explicitly shown that the modified Dirac's equation is separable
in the $D = 5$ EMCS-Kerr-AdS black hole spacetime. In this section, we will demonstrate that
this separability is closely related to the existence of a rank-3 generalized Killing-Yano
tensor admitted by this spacetime metric. To this end, we will show that the separation
constant introduced in the separated radial and angular parts of the modified Dirac equation
acts as the eigenvalue of the first-order differential operator $\mathbb{H}_f$
\be
\big(\mathbb{H}_f +\lambda\big)\Psi = 0 \, ,
\label{dualspe}
\ee
which implies that it commutes with the modified Dirac operator.

This symmetry operator is expressed in terms of the rank-3 generalized Killing-Yano tensor
and explicitly given by Eq. (\ref{dualDiracop}). Using an identity $f^{\rho}_{~\mu\nu;\rho}
= 0$ and the definition $W_{\mu\nu\rho\sigma} = -f_{\mu\nu\rho;\sigma} +f_{\nu\rho\sigma;\mu}
-f_{\rho\sigma\mu;\nu} +f_{\sigma\mu\nu;\rho}$ as well as the anti-commutative property of
gamma matrices, one can also write this operator in another equivalent form
\be
\mathbb{H}_f = -\frac{1}{2}\gamma^{\mu}\gamma^{\nu}f^{~~\rho}_{\mu\nu}
 \big(\p_{\rho} +\Gamma_{\rho} +iq\mathcal{A}_{\rho}\big)
 -\frac{1}{64}\gamma^{\mu}\gamma^{\nu}\gamma^{\rho}\gamma^{\sigma} W_{\mu\nu\rho\sigma} \, .
\ee

We now proceed to write out the explicit expression of this operator in the 5-dimensional
EMCS-Kerr-AdS black hole background. We first compute the term $-\frac{1}{2}\gamma^{\mu}
\gamma^{\nu}f^{~~\rho}_{\mu\nu}\big(\nabla_{\rho} +iq\mathcal{A}_{\rho}\big) = -\frac{1}{2}
\gamma^{\mu} \gamma^{\nu}f^{~~\rho}_{\mu\nu}\big(\p_{\rho} +\Gamma_{\rho} +iq\mathcal{A}_{\rho}
\big)$. After some tedious algebra manipulations, we get
\bea
&& -\frac{1}{2}\gamma^{\mu}\gamma^{\nu}f^{~~\rho}_{\mu\nu} \big(\p_{\rho}
 +\Gamma_{\rho} +iq\mathcal{A}_{\rho}\big) \nn \\
&&\quad = \gamma^5\gamma^0p\sqrt{\frac{\Delta_r}{\Sigma}}\Big(\p_r
 +\frac{\Delta_r^{\prime}}{4\Delta_r} +\frac{1}{2r} +\frac{r}{2\Sigma}\Big)
 +\gamma^5\gamma^1p\frac{1}{r^2\sqrt{\Delta_r\Sigma}}\Big[(r^2+a^2)(r^2+b^2)\p_t \nn \\
&&\qquad +(r^2+b^2)a\chi_a\p_{\phi} +(r^2+a^2)b\chi_b\p_{\psi}
 +Q\big(ab\p_t +b\chi_a\p_{\phi} +a\chi_b\p_{\psi}\big) +\frac{\sqrt{3}}{2}iqQr^2\Big] \nn \\
&&\qquad +\gamma^5\gamma^2(-r)\frac{(p^2-a^2)(p^2-b^2)}{p^2\sqrt{\Delta_p\Sigma}}
 \Big(\p_t -\frac{a\chi_a}{p^2-a^2}\p_{\phi} -\frac{b\chi_b}{p^2-b^2}\p_{\psi}\Big)
 +\gamma^5\gamma^3r\sqrt{\frac{\Delta_p}{\Sigma}}\Big(\p_p
 +\frac{\Delta_p^{\prime}}{4\Delta_p} \nn \\
&&\qquad +\frac{1}{2p} +\frac{p}{2\Sigma}\Big) +\big(p\gamma^0\gamma^1
 -r\gamma^2\gamma^3\big)\frac{1}{rp}\big(ab\p_t +b\chi_a\p_{\phi} +a\chi_b\p_{\psi}\big)
 +\frac{Q}{2\Sigma} -\frac{Q+ab}{2r^2} +\frac{ab}{2p^2} \nn \\
&&\qquad\quad -\Big(\frac{ab}{rp}+\frac{Qp}{r\Sigma}\Big)i\gamma^5
 +i\sqrt{\frac{\Delta_r}{\Sigma}}\Big(\frac{p^2}{2\Sigma}
 -\frac{3}{2}\Big)\gamma^0 +i\sqrt{\frac{\Delta_p}{\Sigma}}\Big(\frac{3}{2}
 -\frac{r^2}{2\Sigma}\Big)\gamma^3 \, .
\label{HFe}
\eea

Next, we consider the last term in the operator $\mathbb{H}_f$. This can be easily done
by considering the exterior differential of the rank-3 generalized Killing-Yano tensor
\be
W = df = -4\Big(\frac{ab}{rp}+\frac{Qp}{r\Sigma}\Big) ~e^0\wedge e^1\wedge e^2\wedge e^3
 -4\sqrt{\frac{\Delta_p}{\Sigma}} ~e^0\wedge e^1\wedge e^2\wedge e^5
 +4\sqrt{\frac{\Delta_r}{\Sigma}} ~e^1\wedge e^2\wedge e^3\wedge e^5 \, ,
\ee
with which we obtain the counter-term
\be
-\frac{1}{64}\gamma^{\mu}\gamma^{\nu}\gamma^{\rho}\gamma^{\sigma} W_{\mu\nu\rho\sigma}
 = \frac{3i}{2}\bigg[\gamma^0\sqrt{\frac{\Delta_r}{\Sigma}}
 -\gamma^3\sqrt{\frac{\Delta_p}{\Sigma}}
 +\gamma^5\Big(\frac{ab}{rp}+\frac{Qp}{r\Sigma}\Big)\bigg] \, ,
\ee
that can exactly cancel the last three unexpected terms in Eq. (\ref{HFe}).

Adding these two terms together, the dual equation (\ref{dualspe}) for the first-order
symmetry operator now reads
\bea
&& \bigg\{ \gamma^5\gamma^0p\sqrt{\frac{\Delta_r}{\Sigma}}\Big(\p_r
 +\frac{\Delta_r^{\prime}}{4\Delta_r} +\frac{1}{2r} +\frac{r -ip \gamma^5}{2\Sigma}\Big)
 +\gamma^5\gamma^1p\frac{1}{r^2\sqrt{\Delta_r\Sigma}}\Big[(r^2+a^2)(r^2+b^2)\p_t \nn \\
&&\qquad +(r^2+b^2)a\chi_a\p_{\phi} +(r^2+a^2)b\chi_b\p_{\psi} +Q\big(ab\p_t
 +b\chi_a\p_{\phi} +a\chi_b\p_{\psi}\big) +\frac{\sqrt{3}}{2}iqQr^2\Big] \nn \\
&&\qquad +\gamma^5\gamma^2(-r)\frac{(p^2-a^2)(p^2-b^2)}{p^2\sqrt{\Delta_p\Sigma}}\Big(\p_t
 -\frac{a\chi_a}{p^2-a^2}\p_{\phi} -\frac{b\chi_b}{p^2-b^2}\p_{\psi}\Big) \nn \\
&&\qquad +\gamma^5\gamma^3r\sqrt{\frac{\Delta_p}{\Sigma}}\Big(\p_p
 +\frac{\Delta_p^{\prime}}{4\Delta_p} +\frac{1}{2p} +\frac{p+ir\gamma^5}{2\Sigma}\Big)
 +\big(p\gamma^0\gamma^1 -r\gamma^2\gamma^3\big)\frac{1}{rp}\big(ab\p_t +b\chi_a\p_{\phi} \nn \\
&&\qquad\quad +a\chi_b\p_{\psi}\big) +\frac{iab}{2rp}\gamma^5 +\frac{iQp}{2r\Sigma}\gamma^5
 +\frac{Q}{2\Sigma} -\frac{Q+ab}{2r^2} +\frac{ab}{2p^2} +\lambda \bigg\}\Psi = 0 \, .
\eea
To decouple this equation, we multiply it a factor $(r-i\gamma^5p)\sqrt{r+i\gamma^5p}\gamma^5$
from the left and arrive at
\bea
&& \bigg\{ \gamma^0p\sqrt{\Delta_r}\Big(\p_r +\frac{\Delta_r^{\prime}}{4\Delta_r}
 +\frac{1}{2r}\Big) +\gamma^1p\frac{1}{r^2\sqrt{\Delta_r}}\Big[(r^2+a^2)(r^2+b^2)\p_t
 +(r^2+b^2)a\chi_a\p_{\phi} \nn \\
&&\qquad +(r^2+a^2)b\chi_b\p_{\psi} +Q\big(ab\p_t +b\chi_a\p_{\phi} +a\chi_b\p_{\psi}\big)
 +\frac{\sqrt{3}}{2}iqQr^2\Big] \nn \\
&&\qquad +\gamma^2(-r)\frac{(p^2-a^2)(p^2-b^2)}{p^2\sqrt{\Delta_p}}\Big(\p_t
 -\frac{a\chi_a}{p^2-a^2}\p_{\phi} -\frac{b\chi_b}{p^2-b^2}
 \p_{\psi}\Big) \nn \\
&&\qquad  +\gamma^3r\sqrt{\Delta_p}\Big(\p_p +\frac{\Delta_p^{\prime}}{4\Delta_p}
 +\frac{1}{2p}\Big) -i\gamma^0\gamma^1\frac{\Sigma}{rp}\big(ab\p_t +b\chi_a\p_{\phi}
 +a\chi_b\p_{\psi}\big) \nn \\
&&\qquad +\frac{i(Q+ab)p}{2r^2} +\frac{abr}{2p^2}\gamma^5
 +\lambda\big(\gamma^5r -ip\big)\bigg\} \big(\sqrt{r +ip\gamma^5}\Psi\big) = 0 \, .
\eea

After using the relation $i\gamma^5 = \gamma^0\gamma^1\gamma^2\gamma^3$, the above equation
can be split as
\bea
&& \bigg\{ \gamma^0\frac{1}{r^2\sqrt{\Delta_r}}\Big[(r^2+a^2)(r^2+b^2)\p_t
 +(r^2+b^2)a\chi_a\p_{\phi} +(r^2+a^2)b\chi_b\p_{\psi} +\frac{\sqrt{3}}{2}iqQr^2 \nn \\
&&\qquad +Q\big(ab\p_t +b\chi_a\p_{\phi} +a\chi_b\p_{\psi}\big)\Big]
 +\gamma^1\sqrt{\Delta_r}\Big(\p_r +\frac{\Delta_r^{\prime}}{4\Delta_r}
 +\frac{1}{2r}\Big) -\frac{i}{r}\big(ab\p_t +b\chi_a\p_{\phi} \nn \\
&&\qquad\quad +a\chi_b\p_{\psi}\big) +\frac{Q+ab}{2r^2}i\gamma^0\gamma^1 +\mu_er
 -i\lambda\gamma^0\gamma^1 \bigg\}\big(\sqrt{r +ip\gamma^5}\Psi\big) = 0 \, , \\
&& \bigg\{ \gamma^2\sqrt{\Delta_p}\Big(\p_p +\frac{\Delta_p^{\prime}}{4\Delta_p}
 +\frac{1}{2p}\Big) +\gamma^3\frac{(p^2-a^2)(p^2-b^2)}{p^2\sqrt{\Delta_p}}\Big(\p_t
 -\frac{a\chi_a}{p^2-a^2}\p_{\phi} -\frac{b\chi_b}{p^2-b^2}\p_{\psi}\Big) \nn \\
&&\qquad +\frac{\gamma^5}{p}\big(ab\p_t +b\chi_a\p_{\phi} +a\chi_b\p_{\psi}\big)
  +\frac{iab}{2p^2}\gamma^0\gamma^1 -i\mu_ep\gamma^5
  +i\lambda\gamma^0\gamma^1 \bigg\}\big(\sqrt{r +ip\gamma^5}\Psi\big) = 0 \, , \qquad
\eea
which reduce to the separated radial and angular equations (\ref{sdea}-\ref{sded}) when
the explicit representations (\ref{GMr}) for gamma matrices have been used.

From this, we can see that $\widetilde{\mathbb{H}}_D\mathbb{H}_f\Psi = \mu_e\lambda\Psi
= \mathbb{H}_f\widetilde{\mathbb{H}}_D\Psi$ and draw a conclusion that it is the existence
of a rank-3 generalized Killing-Yano tensor that ensures the separability of the modified
Dirac equation in the five-dimensional EMCS-Kerr-AdS black hole geometry.

The first-order symmetry operator $\mathbb{H}_f$ can be thought of as the ``square root''
of the second-order operator $\mathbb{K}_c$. It has a lot of correspondences in different
contexts. It is a five-dimensional analogue to the nonstandard Dirac operator discovered
by Carter and McLenaghan \cite{CM} for the four-dimensional Kerr metric. This operator
corresponds to the nongeneric supersymmetric generator in pseudo-classical mechanics
\cite{Spcm}.

\section{Conclusions} \label{CoRe}

In this paper, the hidden symmetries of the general, non-extremal, rotating, charged,
Chong-Cveti\v{c}-L\"{u}-Pope \cite{CCLP} (EMCS-Kerr-AdS) black holes in minimal
five-dimensional gauged supergravity theory have been completely studied. In particular,
we have shown that the existence of the St\"{a}ckel-Killing tensor ensures the separation
of variables in a massive Klein-Gordon complex scalar field equation, and the separability
of a modified Dirac's equation in this spacetime background is also closely associated with
the existence of a generalized Killing-Yano tensor of rank-3.

There are a number of novel characters of this paper. First of all, the whole discussions
of our work are elegant and concise because they are based upon a local Lorentzian
orthonormal pentad that we have set up for the metric. A special advantage of this is
that both the St\"{a}ckel-Killing tensor and the generalized (conformal) Killing-Yano
tensors have simple expressions within the pentad formalism. Next, we have, for the
first time, proposed a suitable generalization of the concepts of (conformal) Killing-Yano
tensors so that they can be subject to minimal $D = 5$ gauged supergravity theory. In our
generalization, there is no need to identify the dual Maxwell three-form with the auxiliary
``torsion'' field. Third, we have constructed two new symmetry operators that, respectively,
commute with the complex scalar Laplacian operator and the modified Dirac operator. Finally,
we have extended thermodynamics to the case of a variable cosmological constant.

There are also many possible applications of the present work, which can serve as a
basis to study various aspects of the spin-$1/2$ spinor field, such as Hawking radiation,
quasinormal modes, absorption rate, perturbation instability, supersymmetry, nongeneric
supersymmetric pseudo-classical mechanics, etc. An interesting and open question is to
investigate whether the present work can be applied or slightly extended to more general
five-dimensional Cveti\v{c}-Youm black holes \cite{CY} with three different charges and
two independent angular momenta.

\smallskip
\textbf{Acknowledgments}:
This work is partially supported by the Natural Science Foundation of China under Grant No.
10675051, 10975058.

\section*{Appendix:~~ Connection one-forms
and Weyl curvature two-forms}

\def\theequation{A\arabic{equation}}
\setcounter{equation}{0}

In this appendix, we outline the computation process for the spin-connection one-forms, the
spinor-connection one-forms, and the curvature two-forms within the f\"{u}nfbein formalism.

Once the pentad coframe one-forms $e^A = e^A_{~\mu}dx^{\mu}$ given in Eq. (\ref{pentad})
have been chosen, we first need to figure out their exterior differentials, and then use
the torsion-free condition --- the Cartan's first structure equation and the skew-symmetric
condition
\be
de^A +\omega^A_{~B}\wedge e^B = 0 \, , \qquad\qquad
\omega_{AB} = \eta_{AC}\omega_{~B}^C = -\omega_{BA} \, ,
\ee
to uniquely determine the spin-connection one-forms $\omega^A_{~B} = \omega^A_{~B\mu}dx^{\mu}
= \Upsilon^A_{~BC}e^C$ in the orthonormal pentad coframe.

In order to obtain the spinor connection one-forms $\Gamma = \Gamma_{\mu}dx^{\mu}\equiv
\Gamma_Ae^A$ from the spin connection one-forms $\omega_{AB} = \omega_{AB\mu}dx^{\mu}
\equiv \Upsilon_{ABC}e^C$, one can utilize the homomorphism between the SO(4,1) group
and its spinor representation derivable from the Clifford algebra (\ref{Clifford}). The
\texttt{so}(4,1) Lie algebra is defined by the ten antisymmetric generators $\Sigma^{AB}
= [\gamma^A, \gamma^B]/(2i)$ which gives the spinor representation, and the spinor connection
$\Gamma$ can be regarded as a \texttt{so}(4,1) Lie-algebra-valued one-form. Using the
isomorphism between the \texttt{so}(4,1) Lie algebra and its spinor representation, i.e.,
$\Gamma_{\mu} = (i/4)\Sigma^{AB}\omega_{AB\mu} = (1/8)[\gamma^A,\gamma^B]\omega_{AB\mu}
= (1/4)\gamma^A \gamma^B\omega_{AB\mu}$, one can immediately construct the spinor connection
one-forms
\be
\Gamma = \frac{1}{8}\big[\gamma^A, \gamma^B\big]\omega_{AB}
= \frac{1}{4}\gamma^A\gamma^B\omega_{AB} = \frac{1}{4}\gamma^A\gamma^B\Upsilon_{ABC}e^C \, ,
\ee
where $\Gamma_A = e_A^{~\mu}\Gamma_{\mu} = (1/4)\gamma^B\gamma^C\Upsilon_{BCA}$ is the
component of the spinor connection in the local Lorentzian frame.

At the last step, we can easily read $\Gamma_A$ from the spinor connection one-forms $\Gamma
\equiv \Gamma_Ae^A = (1/4)\gamma^A\gamma^B\omega_{AB}$. The explicit expressions for the
spin-connection one-forms and the five pentad components of the spinor-connection one-forms
are presented in Eqs. (A2) and (A3) of \cite{WuCY}, where $\Delta_r$ and $\Delta_p$ should
be replaced by the corresponding expressions given in this paper (subject to the case with
a nonzero cosmological constant). We refer the reader to the appendix of our previous paper
\cite{WuCY}.

Taking use of the properties of gamma matrices together with the relation $\gamma^5 =
-i\gamma^0\gamma^1\gamma^2\gamma^3$, we get
\bea
\gamma^A\Gamma_A  &=& \frac{1}{4}\gamma^A\gamma^B\gamma^C\Upsilon_{BCA} \nn \\
&=& \gamma^1\sqrt{\frac{\Delta_r}{\Sigma}}\Big(\frac{\Delta_r^{\prime}}{4\Delta_r}
 +\frac{1}{2r} +\frac{r -ip\gamma^5}{2\Sigma}\Big) +\gamma^2\sqrt{\frac{\Delta_p}{\Sigma}}
 \Big(\frac{\Delta_p^{\prime}}{4\Delta_p} +\frac{1}{2p} +\frac{p
 +ir\gamma^5}{2\Sigma}\Big) \nn \\
&& +\Big(\frac{ab}{2r^2p^2} +\frac{Q}{2r^2\Sigma}\Big)i\gamma^0\gamma^1\big(r +ip\gamma^5\big)
  -\frac{Q}{2\Sigma^2}i\gamma^0\gamma^1\big(r -ip\gamma^5\big) \, ,
\eea
where a prime denotes the partial differential with respect to the coordinates $r$ and $p$.

Using our pentad (\ref{pentad}), the Riemann curvature two-forms $\mathcal{R}^A_{~B}
= d\omega^A_{~B} +\omega^A_{~C}\wedge \omega^C_{~B}$ can be concisely expressed by Eq. (A5)
in the appendix of Ref. \cite{WuCY}, where one has to add a term $-1/l^2$ into the expressions
of the coefficients ($\alpha, \beta, \gamma, \delta, \varepsilon$), while $(C_0, C_1, C_2,
C_3, C_4)$ remain formally unchanged and are given below.

The Ricci tensors, the scalar curvature, and the Einstein tensors for the $D = 5$ EMCS-Kerr-AdS
metric are
\bea
&&\qquad -R_{00} = R_{11} = -\frac{4}{l^2} -\frac{2Q^2(2r^2+p^2)}{\Sigma^4} \, , \quad
 R_{22} = R_{33} = -\frac{4}{l^2} +\frac{2Q^2(r^2+2p^2)}{\Sigma^4} \, , \\
&&\qquad\quad R_{55} = -\frac{4}{l^2} +\frac{2Q^2(r^2-p^2)}{\Sigma^4} \, . \qquad\qquad
 R = -\frac{20}{l^2} -\frac{2Q^2(r^2-p^2)}{\Sigma^4} \, . \\
&& -G_{00} = G_{11} = \frac{6}{l^2} -\frac{3Q^2}{\Sigma^3} \, , \quad
 G_{22} = G_{33} = \frac{6}{l^2} +\frac{3Q^2}{\Sigma^3} \, , \quad
 G_{55} = \frac{6}{l^2} +\frac{3Q^2(r^2-p^2)}{\Sigma^4} \, . \qquad
\eea

Using the pentad (\ref{pentad}), the $U(1)$ gauge potential one-form can be written as
\be
\mathcal{A} = \frac{\sqrt{3}Q}{2\sqrt{\Delta_r\Sigma}} e^0 \, ,
\ee
the field strength two-form and its corresponding Hodge dual three-form are
\bea
F &=& d\mathcal{A} = \frac{\sqrt{3}Q}{\Sigma^2}\big(r ~e^0\wedge e^1
 -p ~e^2\wedge e^3\big) \, , \\
\widetilde{F} &=& {^*}F = \frac{\sqrt{3}Q}{\Sigma^2}\big(p ~e^0\wedge e^1
+r ~e^2\wedge e^3\big)\wedge e^5 \, .
\eea

The complete Einstein equations are satisfied by the energy-momentum tensor of $U(1)$
gauge field
\be
 -T_{00} = T_{11} = -\frac{3Q^2}{2\Sigma^3} \, , \qquad\qquad
 T_{22} = T_{33} = +\frac{3Q^2}{2\Sigma^3} \, , \qquad\qquad
 T_{55} = \frac{3Q^2(r^2-p^2)}{2\Sigma^4} \, .
\ee
The Maxwell-Chern-Simons equation can be rewritten as
\be
\p_{\nu}\big(\sqrt{-g}F^{\mu\nu}\big) +\frac{1}{2\sqrt{3}}
  \epsilon^{\mu\nu\alpha\beta\gamma}F_{\nu\alpha}F_{\beta\gamma} = 0 \, ,
\ee
and is satisfied by verifying that
\be
d\widetilde{F} = -\frac{4\sqrt{3}Q^2rp}{\Sigma^4}~e^0\wedge e^1\wedge e^2\wedge e^3
 = \frac{2}{\sqrt{3}}F\wedge F \, .
\ee

Finally, we present the explicit expression for the Weyl curvature two-forms in the pentad
formalism as follows:
\bea
&& \mathcal{C}^0_{~1} = A ~e^0\wedge e^1 +2C_1 ~e^1\wedge e^5
 -2C_0 ~e^2\wedge e^3 +2C_2 ~e^2\wedge e^5 \, , \nn \\
&& \mathcal{C}^0_{~2} = B ~e^0\wedge e^2 -C_0 ~e^1\wedge e^3
 +C_2 ~e^1\wedge e^5 -C_1 ~e^2\wedge e^5 \, , \nn \\
&& \mathcal{C}^0_{~3} = B ~e^0\wedge e^3 -C_3 ~e^0\wedge e^5
 +C_0 ~e^1\wedge e^2 -C_1 ~e^3\wedge e^5 \, , \nn \\
&& \mathcal{C}^0_{~5} = -C_3 ~e^0\wedge e^3 +C ~e^0\wedge e^5
  -C_2 ~e^1\wedge e^2 \, , \nn \\
&& \mathcal{C}^1_{~2} = -C_0 ~e^0\wedge e^3 +C_2 ~e^0\wedge e^5
 +B ~e^1\wedge e^2 -C_4 ~e^3\wedge e^5 \, , \nn \\
&& \mathcal{C}^1_{~3} = C_0 ~e^0\wedge e^2 +B ~e^1\wedge e^3
 -C_3 ~e^1\wedge e^5 +C_4 ~e^2\wedge e^5 \, , \nn \\
&& \mathcal{C}^1_{~5} = -2C_1 ~e^0\wedge e^1 -C_2 ~e^0\wedge e^2
 -C_3 ~e^1\wedge e^3 +C ~e^1\wedge e^5 +2C_4 ~e^2\wedge e^3 \, , \nn \\
&& \mathcal{C}^2_{~3} = 2C_0 ~e^0\wedge e^1 +2C_4 ~e^1\wedge e^5
 +D ~e^2\wedge e^3 +2C_3 ~e^2\wedge e^5 \, , \nn \\
&& \mathcal{C}^2_{~5} = -2C_2 ~e^0\wedge e^1 +C_1 ~e^0\wedge e^2
 +C_4 ~e^1\wedge e^3 +2C_3 ~e^2\wedge e^3 -C ~e^2\wedge e^5 \, , \nn \\
&& \mathcal{C}^3_{~5} = C_1 ~e^0\wedge e^3 -C_4 ~e^1\wedge e^2
 -C ~e^3\wedge e^5 \, ,
\eea
where
\bea
&& A = \frac{2M}{\Sigma^3}\big(3r^2-p^2\big) -\frac{8Qab}{\Sigma^3}
 -\frac{Q^2(15r^2+11p^2)}{2\Sigma^4} \, , \nn \\
&& B = -\frac{2M}{\Sigma^3}\big(r^2-p^2\big) +\frac{4Qab}{\Sigma^3}
 +\frac{Q^2(5r^2+3p^2)}{2\Sigma^4} \, , \nn \\
&& C = -\frac{2M}{\Sigma^2} +\frac{5Q^2}{2\Sigma^3} \, , \qquad
 D = \frac{2M}{\Sigma^3}\big(r^2-3p^2\big) -\frac{8Qab}{\Sigma^3}
 -\frac{Q^2(5r^2+p^2)}{2\Sigma^4} \, , \nn \\
&& C_0 = \frac{4Mrp}{\Sigma^3} +\frac{2Qab(r^2-p^2)}{rp\Sigma^3}
 -\frac{Q^2(3r^2+2p^2)p}{r\Sigma^4} \, , \nn \\
&& C_1 = \frac{2Qrp}{\Sigma^3}\sqrt{\frac{\Delta_r}{\Sigma}} \, , \quad
 C_2 = -\frac{2Qr^2}{\Sigma^3}\sqrt{\frac{\Delta_p}{\Sigma}} \, , \quad
 C_3 = -\frac{2Qrp}{\Sigma^3}\sqrt{\frac{\Delta_p}{\Sigma}} \, , \quad
 C_4 = \frac{2Qp^2}{\Sigma^3}\sqrt{\frac{\Delta_r}{\Sigma}} \, . \nn
\eea

The non-vanishing components of the Weyl tensor are given by the following one real, two
complex scalars
\bea
&& \frac{1}{4}(A-D) = -\frac{C}{2} = \frac{M}{\Sigma^2}
 -\frac{5Q^2}{4\Sigma^3} \, , \\
&&\Psi_2 = \frac{1}{4}(A+D) +iC_0 = -B +iC_0 \nn \\
&&\quad~ = \frac{2M}{(r-ip)^2\Sigma} -\frac{Q^2(5r-4ip)}{2r(r-ip)^2\Sigma^2}
  +\frac{2iQab}{rp(r-ip)^2\Sigma} \, , \\
&& \frac{C_1+iC_4}{\sqrt{2}}
 = \frac{Qp\sqrt{2\Delta_r}}{(r-ip)\Sigma^{5/2}} \, , \\
&& \frac{C_2+iC_3}{\sqrt{2}}
 = \frac{-Qr\sqrt{2\Delta_p}}{(r-ip)\Sigma^{5/2}} \, ,
\eea
while the only non-zero Maxwell scalar is
\be
F_{01}+iF_{23} = \frac{\sqrt{3}Q}{(r+ip)\Sigma} \, .
\ee
These scalar invariants characterize the properties of the $D = 5$ EMCS-Kerr-AdS spacetime.


\end{document}